\documentclass{jfm}
\usepackage{graphicx}
\usepackage{epstopdf, epsfig}

\shorttitle{Inflation and deflation dynamics of liquid-filled, hyperelastic balloons}
\shortauthor{D. Ilssar and A. D. Gat}

\title{On the inflation and deflation dynamics of liquid-filled, hyperelastic balloons}
\author{Dotan Ilssar and Amir D. Gat}
\affiliation{Faculty of Mechanical Engineering, Technion - Israel Institute of Technology, Haifa 3200003, Israel}

\begin{document}

\maketitle

\begin{abstract}
We derive a reduced-order model describing the inflation and deflation dynamics of a liquid-filled hyperelastic balloon, focusing on inviscid laminar flow and the extensional motion of the balloon. We initially study the flow and pressure fields for dictated motion of the solid, which throughout deflation are obtained by solving the potential problem. However, during inflation, flow separation creates a jet within the balloon, requiring a different approach. The analyses of both flow regimes lead to a simple piecewise model, describing the fluidic pressure during inflation and deflation, which is then verified by finite element computations. We then use a variational approach to derive the equation governing the balloon's dynamics, yielding a nonlinear hybrid oscillator equation, describing the interaction between the extensional mode of the balloon, and the entrapped fluid. Analytical and graphical investigations of the suggested model are presented, shedding light on its static and dynamic behaviour under different operating conditions. Our suggested model and its underlying assumptions are verified utilizing a fully coupled finite element scheme, showing excellent agreement.
\end{abstract}

\begin{keywords}

\end{keywords}

\section{Introduction}
Balloons are a key engineering element as they are used for medical applications, such as pleural pressure assessments \citep[][]{milic1964improved,yang2017optimal} and enteroscopy \citep[][]{kawamura2008clinical,yamamoto2001total}, for aviation purposes with the example of super pressure balloons \citep[][]{cathey2009nasa,saito2014development}, for meteorological measurements \citep[][]{sankar2009embedded}, and more. Recent studies in the field of soft robotics use elastic materials, embedded with internal cavities, whose motion is governed by the controlled pressure inside these cavities \citep[][]{gamus2017interaction,mosadegh2014pneumatic,siefert2019bio}. Namely, the embedded cavities expand and contract due to pressure variations in the entrapped fluid, deforming the solid to a designated state, while careful switching between these different states lead to controlled locomotion. 

A formerly suggested configuration of a self-propelled soft robot consists of an array of serially connected spherical hyperelastic structures such as rubber balloons, which exhibit bi-stability, meaning that at a certain pressure range, they have two stable equilibrium radii \citep[][]{ben2019single,glozman2010self}. Thanks to their bi-stability property, interconnection of these spherical balloons composes a multi-stable system, where each stable equilibrium corresponds to a pattern in which each balloon is inflated to one of its two stable equilibria. As shown by the abovementioned studies, the equilibrium state of the multi-stable robot can be chosen utilizing a single pressure input, since the pressure inside the different balloons does not vary instantly when changing the pressure at the inlet. Instead, the pressure variation starts at the element connected to the inlet, and progress, allowing to selectively inflate each balloon to one of its stable states, by varying the pressure at the input according to a certain carefully synthesized profile. Therefore, embedding spherical cavities inside hyperelastic structures can pave the way toward manufacturing soft robots which utilize minimal actuation points to produce highly complex walking patterns. However, in order to assess the dynamics of such robots under applicable time constants, a complete mathematical model, describing their fully coupled, fluid-structure interaction at high Reynolds numbers, is needed.

The behaviour of spherical hyperelastic balloons is the subject of numerous studies, examining the validity of different hyperelastic models, the stability of their spherical shape, the behaviour of interconnected balloons, etc. \citep[][]{ben2019single,mangan2015gent,muller2004rubber,wang2018snap}. However, all of these studies consider a uniform pressure field inside the balloons, describing a quasi-static behaviour. Moreover, most analytical studies, dealing with flow inside a spherical cavity, consider a creeping flow, implying on very low Reynolds numbers \citep[][]{maul1994image,usha1993flow}. The only exception found in the literature, is an article by \cite{wang1979dynamic}, providing a model of the fluid-field inside a spherical cavity with time-varying boundaries, based on the potential flow theory. As shown below, the latter is a partial model since it does not describe the fluid's behaviour throughout inflation, where boundary-layer separation affects the validity of the assumptions on which the potential flow theory is based. It should be pointed out that all the studies mentioned above, examining the behaviour of hyperelastic balloons and the entrapped fluid, completely disregard the dynamic interaction between them.

The present paper deals with reduced order modelling, describing the fully coupled fluid-structure interaction between a single spherical hyperelastic balloon and the fluid filling it, assuming incompressible inviscid flow.

\section{Reduced order modelling}
\begin{figure}
\centering
\includegraphics[width=0.7\textwidth]{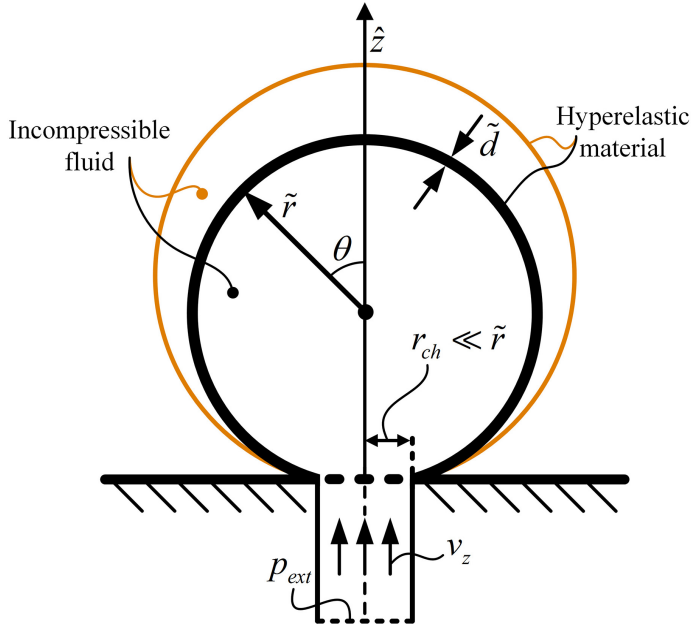}
\caption{Schematic layout of the system}
\label{Figure1}
\end{figure}

We examine the coupled dynamics of a spherical hyperelastic balloon, connected through a rigid channel to an inlet, supplying incompressible inviscid fluid (see figure \ref{Figure1}). The pressure at the inlet, denoted ${p_{ext}}$, is externally controlled and the balloon is assumed to exhibit only extensional motion. Namely, the theoretical system has a single degree of freedom, represented here by the instantaneous average radius of the balloon, denoted $\tilde r$. The modes disregarded in this paper do not have a spherical symmetry, thus they distort the theoretical behaviours of both the fluid and the solid, what can cause deviations between a realistic system and its theoretical form discussed here. However, the validity of this simplifying assumption is examined and justified by numerical computations.

\begin{figure}
\centering
\includegraphics[width=0.4\textwidth]{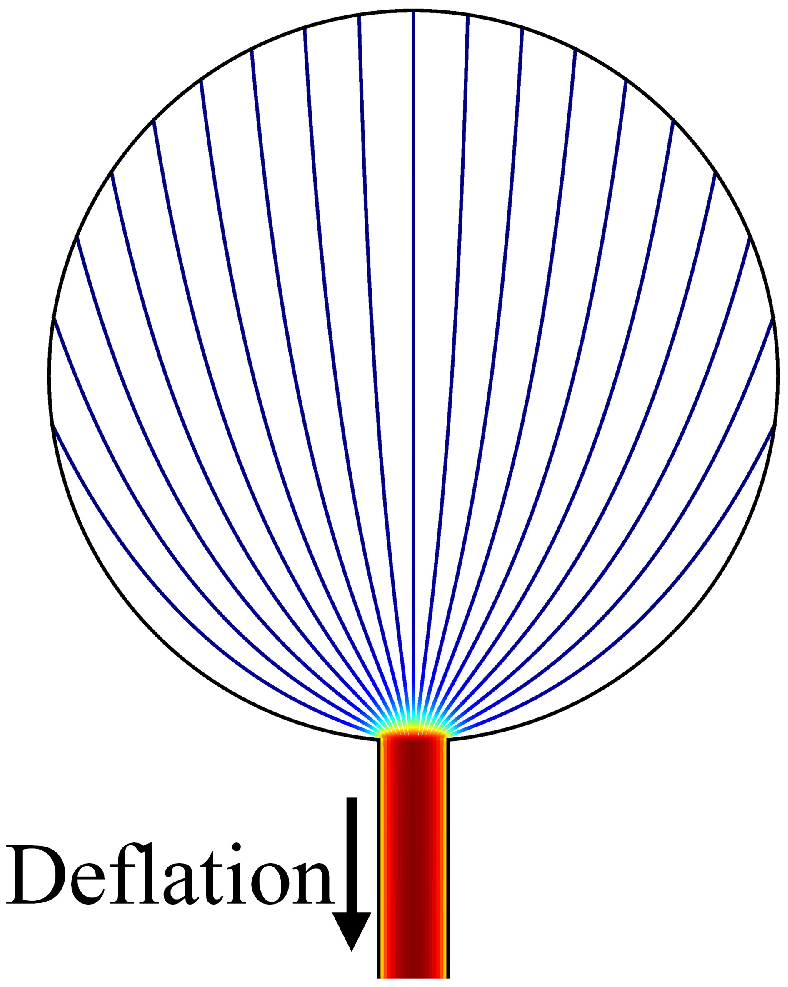}
\qquad
\includegraphics[width=0.4\textwidth]{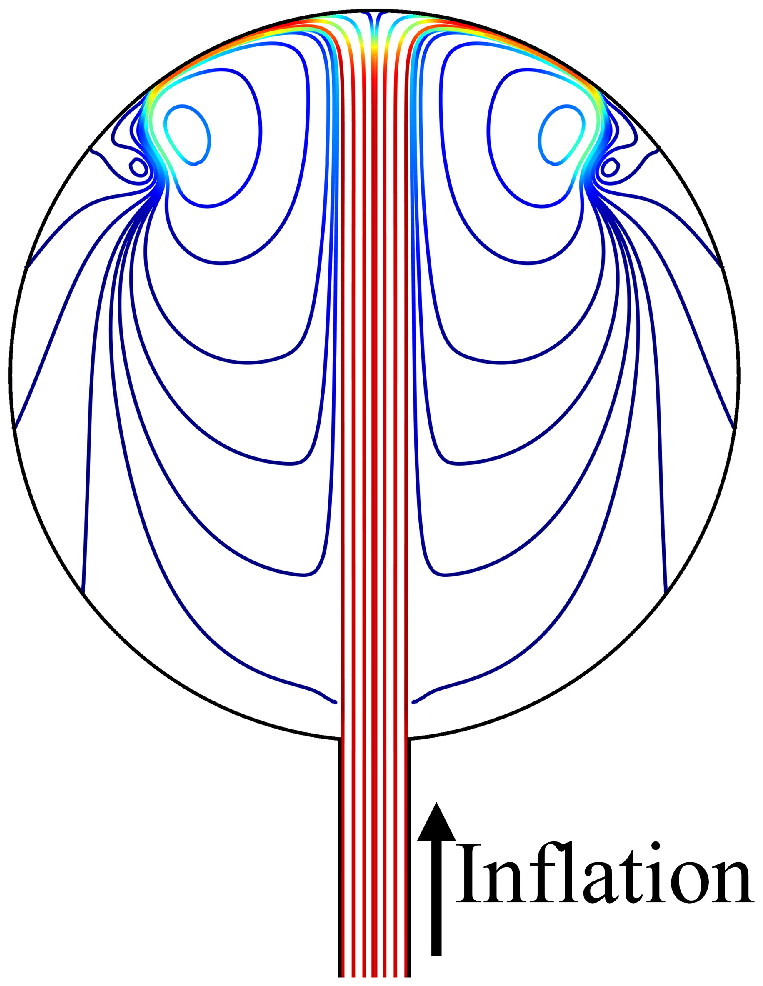}
\caption{Streamlines, describing the typical flow velocity field when the radius of the balloon decreases (left) and increases (right), imitating deflation and inflation respectively, based on finite element simulations carried out utilizing COMSOL Multiphysics. The red regions correspond to the highest velocities, whereas the blue regions represent the lowest velocities.}
\label{Figure2}
\end{figure}

The reduced order model, describing the behaviour of the system in figure \ref{Figure1}, is derived following several steps. First, the behaviour of the entrapped fluid is modelled referring to the motion of the balloon as dictated, while considering the different flow regimes throughout deflation and inflation, see figure \ref{Figure2}. In the former case, under the assumptions of incompressible laminar flow at high Reynolds numbers, irrotationality, and no boundary layer separation, the behaviour of the flow inside both the balloon and the channel are modelled based on the potential flow theory. However, modelling the fluid's behaviour throughout inflation requires a different approach since in this case boundary layer separation causes an internal jet (see figure \ref{Figure2} right). Next, a simplified expression, describing the generalized forces applied by the entrapped fluid on the extensional mode of the balloon, is derived and empirically modified based on finite element analyses, disregarding the dynamics of the balloon. Combining the generalized force applied by the entrapped fluid and a variational model of the hyperelastic balloon, yields a complete reduced-order model, describing the fully coupled dynamics of the system in figure \ref{Figure1}. Finally, some limiting cases of the reduced-order model are compared to finite element simulations of the fully coupled system, for validation.

The derivation of the model, describing the dynamics of the system in figure \ref{Figure1} starts with formulation of the flow field inside the channel, connecting the controlled pressure inlet to the spherical balloon. Here, since the channel is assumed to be straight, and with a constant radius, it should cause no boundary layer separation throughout inflation, nor deflation. Thus, the behaviour of the flow inside the channel should not change between these two regimes, meaning that the upcoming analysis, dealing with the flow field inside the channel, suits both inflation and deflation.

Assuming the system is axisymmetric around the axis of the channel eliminates all dependencies on the tangential coordinate $\varphi$. An additional assumption, requiring that the radius of the channel, denoted ${r_{ch}}$, is considerably smaller compared with its length ${L_{ch}}$, allows referring to the pressure field in every cross-section along the channel as nearly uniform. The latter also eliminates the radial component of the flow velocity field, meaning that inside the channel, the fluid is assumed to be flowing only along the axial coordinate $z$. Thus, together with the incompressibility assumption, the continuity equation in this case is degenerated into \({{\partial {v_z}} \mathord{\left/
 {\vphantom {{\partial {v_z}} {\partial z}}} \right.
 \kern-\nulldelimiterspace} {\partial z}} = 0,\) suggesting that the flow velocity along the axial direction of the channel, denoted ${v_z}$, can vary only in the radial direction. Finally, considering uniform flow velocity profile at the inlet, significantly simplifies the flow field inside the channel since in this case the latter becomes spatially uniform all along the channel. This uniform, axial flow velocity is formulated by dividing the volumetric flow rate by the cross sectional area of the channel, where due to incompressibility, the flow rate is given by the time derivative of the balloon's volume, without the portion penetrating into the channel. Thus, the uniform axial flow velocity inside the channel is:
  \begin{equation}\label{eq21}
 {v_z} = \frac{1}{{\pi r_{ch}^2}}\frac{{\rm{d}}}{{{\rm{d}}t}}\left( {\frac{{4\pi {{\tilde r}^3}}}{3} - \int\limits_0^{2\pi } {\int\limits_{\pi  - {{\sin }^{ - 1}}\left( {{{{r_{ch}}} \mathord{\left/
 {\vphantom {{{r_{ch}}} {\tilde r}}} \right.
 \kern-\nulldelimiterspace} {\tilde r}}} \right)}^\pi  {\int\limits_{{{\sqrt {{{\tilde r}^2} - r_{ch}^2} } \mathord{\left/
 {\vphantom {{\sqrt {{{\tilde r}^2} - r_{ch}^2} } {\cos \left( {\pi  - \theta } \right)}}} \right.
 \kern-\nulldelimiterspace} {\cos \left( {\pi  - \theta } \right)}}}^{\tilde r} {{r^2}\sin \theta {\rm{d}}r{\rm{d}}\theta {\rm{d}}\varphi } } } } \right) = \frac{{2 - \varepsilon }}{{\varepsilon \left( {1 - \varepsilon } \right)}}\frac{{{\rm{d}}\tilde r}}{{{\rm{d}}t}}.
\end{equation}
Here, $r,\theta ,\varphi $ are the coordinates of a non-inertial spherical system, located at the centre of the balloon, where $\theta $ is the angle, measured from the axis of symmetry to the radial coordinate $r$, see figure \ref{Figure1}, and $\varphi$ is the angle, revolving around the axis of symmetry, in similar to the cylindrical system discussed above. Moreover, $\varepsilon  = 1 - \left( {{{\sqrt {{{\tilde r}^2} - r_{ch}^2} } \mathord{\left/
 {\vphantom {{\sqrt {{{\tilde r}^2} - r_{ch}^2} } {\tilde r}}} \right.
 \kern-\nulldelimiterspace} {\tilde r}}} \right) \ll 1$, serving as the small parameter of the system during the model derivation, denotes the ratio between the part of the balloon's radius, penetrating into the channel at $\theta  = \pi $, and the total radius of the balloon.

In order to describe the pressure field inside the balloon in the upcoming sections, the pressure at the imaginary plane located at $z = {L_{ch}}$, serving as the entrance to the balloon, is needed as a boundary condition. For this sake, assuming the fluid behaves according to the potential flow theory, the unsteady Bernoulli equation, which after neglecting gravitational effects is given by \citep[][]{white1994fluid}
  \begin{equation}\label{eq22}
p + {\rho _f}\frac{{\partial \phi }}{{\partial t}} + \frac{{{\rho _f}{{\left| {\nabla \phi } \right|}^2}}}{2} = {p_{stag}}\left( t \right),
\end{equation}
is utilized. This equation relates between the pressure field denoted $p$, the fluid's constant density \({\rho _f}\), and the velocity potential function $\phi$, which is connected to the flow velocity field ${\bf{v}}$ by ${\bf{v}} = \nabla \phi$ , constraining the flow to be irrotational since \(\nabla  \times \nabla \phi  \equiv 0\). In (\ref{eq22}), the right hand side, represented by ${p_{stag}}$ is the stagnation pressure which is an unknown temporal function, which due to irrotationality holds at each point inside the medium, and not only along a streamline \citep[][]{white1994fluid}.

Comparing the stagnation pressure at both sides of the channel after substituting (\ref{eq21}), the relation between ${\bf{v}}$ and $\phi$, and the dictated pressure at the inlet, into (\ref{eq22}), leads to the following formulation, describing the pressure at the connection between the channel and the balloon:
\begin{equation}\label{eq23}
p\left( {z = {L_{ch}},t} \right) = {p_{ext}} - \frac{{{\rho _f}{L_{ch}}\left( {2 - \varepsilon } \right)}}{{\varepsilon \left( {1 - \varepsilon } \right)}}\left[ {\frac{{{{\rm{d}}^2}\tilde r}}{{{\rm{d}}{t^2}}} + \frac{{2 - 4\varepsilon  + {\varepsilon ^2}}}{{\tilde r{{\left( {1 - \varepsilon } \right)}^2}}}{{\left( {\frac{{{\rm{d}}\tilde r}}{{{\rm{d}}t}}} \right)}^2}} \right].
\end{equation}
This formulation alongside the expression given in (\ref{eq21}) serve as boundary conditions to the pressure and flow velocity fields inside balloon at the upcoming sections.
\subsection{Modelling the flow inside the balloon throughout deflation}
As mentioned above, due to the difference between the behaviour of the fluid throughout inflation and deflation, the flow fields corresponding to the different regimes are derived separately, starting with deflation. 

Under the assumptions underlined above, the flow inside the balloon during deflation is assumed irrotational and inviscid, allowing to define a velocity potential function denoted $\phi$, whose relation to the flow velocity field ${\bf{v}}$ is defined above. Further, since the fluid is assumed incompressible, substitution of the relation between ${\bf{v}}$ and $\phi$ into the continuity equation, leads to Laplace equation in terms of the velocity potential function. Here, due to the break of symmetry, caused by the connection to the channel, the flow does not have a spherical symmetry, so it depends on both the radial direction $r$ and the tangential direction $\theta$, of the spherical coordinate system located at the centre of the balloon. However, the balloon is still symmetrical around the $z$ axis, eliminating the dependency on $\varphi$, what degenerates the governing equation into the following form:
\begin{equation}\label{eq24}
\frac{\partial }{{\partial r}}\left( {{r^2}\frac{{\partial \phi }}{{\partial r}}} \right) + \frac{1}{{\sin \theta }}\frac{\partial }{{\partial \theta }}\left( {\sin \theta \frac{{\partial \phi }}{{\partial \theta }}} \right) = 0.
\end{equation}

In order to formulate an expression, describing the flow velocity field inside the balloon throughout deflation, in the upcoming analysis, (\ref{eq24}) is solved together with the appropriate boundary conditions. These consider no penetration into the balloon, and the uniform flow at the connection between the balloon and the channel, given by (\ref{eq21}). Since the model derivation is carried out in the non-inertial spherical coordinate system, the boundary conditions are formulated in terms of this system as well. Moreover, as shown below, since the potential flow theory is applied, taking only the radial component of the boundary conditions is sufficient to provide a closed form explicit expression, describing the flow velocity field. Thus, the boundary conditions derived below deals merely the radial component of the flow velocity. Finally, for simplicity the boundary condition describing the flow velocity at the connection to the channel is taken at the imaginary continuation of the balloon inside the channel, instead of on the imaginary plane, connecting the channel and the balloon. The latter can cause deviations in the flow velocity field inside the volume trapped between these imaginary surfaces, compared to the formulation in (\ref{eq21}). Nevertheless, under the assumption that $\varepsilon  \ll 1$ this volume is insignificant, and the errors caused by the resulting deviations can be neglected.

The first boundary condition, requiring that there is no penetration into the balloon, which is defined at \(0 \le \theta  \le \pi  - {\sin ^{ - 1}}\left( {{{{r_{ch}}} \mathord{\left/
 {\vphantom {{{r_{ch}}} {\tilde r}}} \right.
 \kern-\nulldelimiterspace} {\tilde r}}} \right)\) where $r = \tilde r$, is simply given by \({{\partial \phi } \mathord{\left/
 {\vphantom {{\partial \phi } {\partial r}}} \right.
 \kern-\nulldelimiterspace} {\partial r}} = {{{\rm{d}}\tilde r} \mathord{\left/
 {\vphantom {{{\rm{d}}\tilde r} {{\rm{d}}t}}} \right.
 \kern-\nulldelimiterspace} {{\rm{d}}t}}\). However, thanks to the motion of the spherical coordinate system, to derive the boundary condition at the connection to the channel, defined where \(\pi  - {\sin ^{ - 1}}\left( {{{{r_{ch}}} \mathord{\left/
 {\vphantom {{{r_{ch}}} {\tilde r}}} \right.
 \kern-\nulldelimiterspace} {\tilde r}}} \right) < \theta  \le \pi \), the velocity of the centre of the balloon, given by \({{{\rm{d}}{{\bf{r}}_{\bf{c}}}} \mathord{\left/
 {\vphantom {{{\rm{d}}{{\bf{r}}_{\bf{c}}}} {{\rm{d}}t}}} \right.
 \kern-\nulldelimiterspace} {{\rm{d}}t}} = \left( {{{\tilde r\left( {{{{\rm{d}}\tilde r} \mathord{\left/
 {\vphantom {{{\rm{d}}\tilde r} {{\rm{d}}t}}} \right.
 \kern-\nulldelimiterspace} {{\rm{d}}t}}} \right)} \mathord{\left/
 {\vphantom {{\tilde r\left( {{{{\rm{d}}\tilde r} \mathord{\left/
 {\vphantom {{{\rm{d}}\tilde r} {{\rm{d}}t}}} \right.
 \kern-\nulldelimiterspace} {{\rm{d}}t}}} \right)} {\sqrt {{{\tilde r}^2} - r_{ch}^2} }}} \right.
 \kern-\nulldelimiterspace} {\sqrt {{{\tilde r}^2} - r_{ch}^2} }}} \right)\hat z\), should be subtracted from the flow velocity at the inlet (\ref{eq21}). Thus, the radial flow velocity at the connection to the channel is calculated by projecting the subtraction of \({{{\rm{d}}{{\bf{r}}_{\bf{c}}}} \mathord{\left/
 {\vphantom {{{\rm{d}}{{\bf{r}}_{\bf{c}}}} {{\rm{d}}t}}} \right.
 \kern-\nulldelimiterspace} {{\rm{d}}t}}\) from (\ref{eq21}), on the radial coordinate. Combining the boundary conditions achieved for both regions, and substituting the flow velocity at the inlet given by (\ref{eq21}), and the relation between ${r_{ch}}$ to $\varepsilon$, yields the following, final form of the flow velocity field's boundary conditions:
\begin{equation}\label{eq25}
{\left. {\frac{{\partial \phi }}{{\partial r}}} \right|_{r = \tilde r}} = \frac{{{\rm{d}}\tilde r}}{{{\rm{d}}t}}\left\{ {\begin{array}{*{20}{c}}
1&{0 \le \theta  \le \pi  - {{\sin }^{ - 1}}\sqrt {\varepsilon \left( {2 - \varepsilon } \right)} }\\
{2{\varepsilon ^{ - 1}}\cos \theta }&{\pi  - {{\sin }^{ - 1}}\sqrt {\varepsilon \left( {2 - \varepsilon } \right)}  < \theta  \le \pi .}
\end{array}} \right.
\end{equation}

In the following sections, Laplace equation (\ref{eq24}) together with the boundary conditions given by (\ref{eq25}), provide a closed form solution for the velocity potential function, from which the flow velocity field can be easily extracted. Next, in order to enable coupling the effect of the fluid field, with the dynamic behaviour of the balloon, the pressure distribution is formulated as well. For this sake, in similar to the previous section, the pressure field is derived utilizing the unsteady Bernoulli equation (\ref{eq22}), while considering the pressure at the connection to the channel, given by (\ref{eq23}).

\subsubsection{Formulating the flow velocity field}
In order to formulate the velocity potential function, describing the flow velocity field inside the balloon throughout deflation, the separation of variables solution \(\phi \left( {r,\theta ,t} \right) = R\left( {r,t} \right)\Theta \left( {\cos \theta ,t} \right)\) is suggested. The latter converts the partial differential equation (\ref{eq24}) into radial and tangential differential equations, which can be solved while ignoring the time-dependency. Enforcing the solution to be bounded leads to Eigenfunctions and Eigenvalues whose substitution into the separation of variables solution yields the following general form of the velocity potential function:
\begin{equation}\label{eq26}
\phi \left( {r,\theta ,t} \right) = \frac{{{\rm{d}}\tilde r}}{{{\rm{d}}t}}\sum\limits_{m = 1}^\infty  {\frac{{{A_m}\left( {\varepsilon \left( t \right)} \right){r^m}{P_m}\left( {\cos \theta } \right)}}{{2m{{\tilde r}^{m - 1}}}}} ,
\end{equation}
where \({P_m}\left(  \bullet  \right)\) is a Legendre polynomial of order $m$. Imposing the boundary conditions in (\ref{eq25}), after development into a generalized Fourier series of the tangential Eigenfunctions provides the following closed form expression of the coefficients ${A_m}$:
\begin{equation}\label{eq27}
{A_m} = \left\{ {\begin{array}{*{20}{c}}
{\frac{{12 - 6\varepsilon  + {\varepsilon ^2}}}{2}}&{m = 1}\\
{\frac{2}{\varepsilon }\left[ \begin{array}{l}
\frac{{{P_m}\left( {\varepsilon  - 1} \right) - m\left( {\varepsilon  - 1} \right){P_{m - 1}}\left( {\varepsilon  - 1} \right)}}{{m - 1}}\\
 + \frac{{{P_m}\left( {\varepsilon  - 1} \right) + \left( {m + 1} \right)\left( {\varepsilon  - 1} \right){P_{m + 1}}\left( {\varepsilon  - 1} \right)}}{{m + 2}}
\end{array} \right] + \left[ \begin{array}{l}
{P_{m - 1}}\left( {\varepsilon  - 1} \right)\\
 - {P_{m + 1}}\left( {\varepsilon  - 1} \right)
\end{array} \right]}&{m = 2,3, \cdots .}
\end{array}} \right.
\end{equation}

\subsubsection{Formulating the pressure field and its influence on the balloon} \label{sec212}

As mentioned above, to derive a model describing the fully coupled dynamics of the system in figure \ref{Figure1}, it is necessary to formulate the forces applied by the entrapped fluid on the balloon. For this sake, first the pressure field inside the balloon is derived, utilizing the unsteady Bernoulli equation (\ref{eq22}). Here, since  the stagnation pressure ${p_{stag}}$ is uniform in every instant, it is determined arbitrarily by limiting the expression obtained by substituting (\ref{eq26}) into (\ref{eq22}), to the middle of the connection between the channel and the balloon. I.e. the stagnation pressure is determined at \(\left( {r,\,\theta } \right) \to \left( {\sqrt {{{\tilde r}^2} - r_{ch}^2} ,\,\pi } \right)\), utilizing (\ref{eq23}), quantifying the pressure at the connection, based on the analysis of the flow inside the channel. The unsteady Bernoulli equation, supplemented by the abovementioned formulation of the stagnation pressure, yields the expression describing the pressure field inside the balloon, given in appendix \ref{appA}. It should be noted that Bernoulli equation holds only in a fixed reference frame \citep[][]{mungan2011bernoulli}. Therefore, (\ref{eq22}) was formulated in terms of a fixed coordinate system, whose origin is located at the middle of the connection between the channel and the balloon, followed by returning to the original, non-inertial spherical system. The formulation at the fixed coordinate system, required adding an additional term to the velocity potential function given by \(z{\left( {1 - \varepsilon } \right)^{ - 1}}{{{\rm{d}}\tilde r} \mathord{\left/
 {\vphantom {{{\rm{d}}\tilde r} {{\rm{d}}t}}} \right.
 \kern-\nulldelimiterspace} {{\rm{d}}t}}\), originating in the motion of the spherical coordinate system.

Finally, to describe the effect of the flow field on the balloon, for sake of coupling between them, the generalized force applied by the pressure on the balloon's extensional mode is calculated by surface integration of the pressure field over the balloon. This generalized force, describing the averaged pressure on the balloon multiplied by its surface area, is given by
\begin{eqnarray}\label{eq28}
{F_{deflation}} = \mathop{{\int\!\!\!\!\!\int}\mkern-21mu \bigcirc}\limits_S 
 {p\left( {\tilde r,\theta ,t} \right){\rm{d}}S}  = {g_1}\left( \varepsilon  \right){\tilde r^2}{p_{ext}} & - & {\rho _f}\left[ {{L_{ch}}{g_2}\left( \varepsilon  \right) + {g_3}\left( \varepsilon  \right)\tilde r} \right]{\tilde r^2}\frac{{{{\rm{d}}^2}\tilde r}}{{{\rm{d}}{t^2}}}\nonumber\\
 & + & {\rho _f}\left[ {{L_{ch}}{g_4}\left( \varepsilon  \right) + {g_5}\left( \varepsilon  \right)\tilde r} \right]\tilde r{\left( {\frac{{{\rm{d}}\tilde r}}{{{\rm{d}}t}}} \right)^2},
\end{eqnarray}
where $S$ is the surface area of the balloon, and the functions ${g_1}\left( \varepsilon  \right) \div {g_5}\left( \varepsilon  \right)$ are given in appendix \ref{appB}.

\subsubsection{Simplifying the generalized force} 
The expression describing the generalized force applied by the pressure field on the extensional mode of the balloon, given by (\ref{eq27}), (\ref{eq28}),  (\ref{eqA2}) and (\ref{eqB}), is quite complex, and thus fails to provide physical insight on the system. Thus, in the current section, \({F_{deflation}}\) is simplified  based on the assumption, requiring that $\varepsilon\ll 1$, or equivalently that ${r_{ch}} \ll \tilde r$. 

First, due to their relatively simple forms, the functions ${g_1},\,\,{g_2},\,\,{g_4}$ are approximated analytically. This is done by developing these functions into power series around $\varepsilon  = 0$, keeping only their leading order terms, what yield the following approximations:
\refstepcounter{equation} \label{eq29}
$$
{g_1}\left( \varepsilon  \right) = 4\pi  + O\left( \varepsilon  \right),\,\,\,\,\,\,{g_2}\left( \varepsilon  \right) = 8\pi {\varepsilon ^{ - 1}} + O\left( \varepsilon  \right),\,\,\,\,\,\,{g_4}\left( \varepsilon  \right) =  - 16\pi {\varepsilon ^{ - 1}} + O\left( \varepsilon  \right).
\eqno{(\theequation{\mathit{a},\mathit{b},\mathit{c}})}
$$
Comparison between the exact values of  ${g_1},\,\,{g_2},\,\,{g_4}$, achieved from (\ref{eqB}), and their approximated values calculated utilizing (\ref{eq29}), shows minor deviations which do not exceed 0.3\% if ${{{r_{ch}}} \mathord{\left/
 {\vphantom {{{r_{ch}}} {\tilde r}}} \right.
 \kern-\nulldelimiterspace} {\tilde r}} \le {1 \mathord{\left/
 {\vphantom {1 {10}}} \right.
 \kern-\nulldelimiterspace} {10}}$.
 
 Next, the functions ${g_3},\,\,{g_5}$ which cannot be approximated analytically since they contain infinite series, are estimated numerically. For this, each function is calculated for several values of $\varepsilon$, corresponding to the designated working range, chosen to be \({{{1 \mathord{\left/
 {\vphantom {1 {200}}} \right.
 \kern-\nulldelimiterspace} {200}} \le {r_{ch}}} \mathord{\left/
 {\vphantom {{{1 \mathord{\left/
 {\vphantom {1 {200}}} \right.
 \kern-\nulldelimiterspace} {200}} \le {r_{ch}}} {\tilde r}}} \right.
 \kern-\nulldelimiterspace} {\tilde r}} \le {1 \mathord{\left/
 {\vphantom {1 {10}}} \right.
 \kern-\nulldelimiterspace} {10}}\), where each computation considers the first $2 \times {10^4}$ harmonics in each series. Curve fittings on the numerical data, obtained for the different values of $\varepsilon $, yield the following closed form approximated expressions for ${g_3},\,\,{g_5}$:
  \refstepcounter{equation} \label{eq210}
$$
{g_3}\left( \varepsilon  \right) \approx 36{\varepsilon ^{ - {1 \mathord{\left/
 {\vphantom {1 2}} \right.
 \kern-\nulldelimiterspace} 2}}},\,\,\,\,\,\,{g_5}\left( \varepsilon  \right) \approx 24{\varepsilon ^{ - 2}}.
\eqno{(\theequation{\mathit{a},\mathit{b}})}
$$
The approximated values obtained from (\ref{eq210}), show maximal relative errors of 1\% and 5\% respectively at the working range, compared to those, achieved from (\ref{eqB}) under the abovementioned series truncation. In should be noted that the numerical computations discussed above, disregard the last term in ${g_5}$, containing an integral which increases the computation time, substantially. The omission of this term is justified by comparison between the exact values of ${g_5}$, and the values obtained while dropping the abovementioned term, which shows negligible deviations when summing enough harmonics.

Thanks to the good agreement between the values achieved from the original and the approximated expressions of  ${g_1} \div {g_5}$, it seems reasonable to utilize (\ref{eq29}) and (\ref{eq210}), converting the generalized force into a simplified form, which still depends on the small parameter $\varepsilon $. In order to further simplify this expression, and eliminate the dependency on $\varepsilon $, the latter is developed into a Taylor series around ${{{r_{ch}}} \mathord{\left/
 {\vphantom {{{r_{ch}}} {\tilde r}}} \right.
 \kern-\nulldelimiterspace} {\tilde r}} = 0$. Namely, $\varepsilon  = {{r_{ch}^2} \mathord{\left/
 {\vphantom {{r_{ch}^2} {2{{\tilde r}^2}}}} \right.
 \kern-\nulldelimiterspace} {2{{\tilde r}^2}}} + O\left( {{{{{r_{ch}^4} \mathord{\left/
 {\vphantom {{r_{ch}^4} {\tilde r}}} \right.
 \kern-\nulldelimiterspace} {\tilde r}}}^4}} \right)$ is substituted into the approximated expression of the generalized force, yielding the following convenient form: 
\begin{equation}\label{eq211}
{F_{deflation}} \approx 4\pi {\tilde r^2}{p_{ext}} - \frac{{4{\rho _f}}}{{{r_{ch}}}}\left[ {\frac{{4\pi {L_{ch}}}}{{{r_{ch}}}} + 9\sqrt 2 } \right]{\tilde r^4}\frac{{{{\rm{d}}^2}\tilde r}}{{{\rm{d}}{t^2}}} - \frac{{32{\rho _f}}}{{r_{ch}^2}}\left[ {\pi {L_{ch}} - \frac{{3{{\tilde r}^3}}}{{r_{ch}^2}}} \right]{\tilde r^3}{\left( {\frac{{{\rm{d}}\tilde r}}{{{\rm{d}}t}}} \right)^2}.
\end{equation}

It can be seen that ${F_{deflation}}$ consists five terms, arising from different effects. The first term originates in the dictated, external pressure; the second and fourth terms are the inertial and centripetal forces induced due to the flow inside the channel, and the third and fifth terms are the inertial and centripetal forces exerted by the flow inside the balloon.

\subsection{Modelling the flow inside the balloon throughout inflation}
As shown in figure \ref{Figure2}, throughout inflation, boundary layer separation causes an internal jet inside the balloon, where even under the assumption that the jet is stable, the transient flow velocity field is extremely complex. However, assuming the jet develops into its steady form rapidly, and in similar, its decays rapidly when the inflation stops, it can be modelled as a steady jet. Furthermore, based on the assumption, requiring that the radius of the channel is substantially smaller than the balloon's dimensions, the entrapped medium can be approximated as semi-infinite. Consequentially, well-known results, describing the spread of an unbounded jet can be applied \citep[][]{goldstein1965modern,schlichting2016boundary}. A basic assumption, on which these results are based is that the pressure field in the entire medium is uniform. Thus, in the case considered here, throughout inflation, the pressure is approximated to be completely uniform inside the balloon, where its value is taken as the pressure at the entrance to the balloon, given by (\ref{eq23}). The latter degenerates the generalized force applied on the balloon's extensional mode compared to (\ref{eq211}), so it includes only the terms, originating in the external pressure and the flow inside the channel. 

The complete form of the generalized force, describing both regimes and denoted ${F_p}$, is achieved by combining the formulations discussed above. Namely, ${F_p}$ is taken as \({F_{deflation}}\) when \({{{\rm{d}}\tilde r} \mathord{\left/
 {\vphantom {{{\rm{d}}\tilde r} {{\rm{d}}t}}} \right.
 \kern-\nulldelimiterspace} {{\rm{d}}t}} < 0\), whereas its  degenerated form corresponding to inflation is employed when \({{{\rm{d}}\tilde r} \mathord{\left/
 {\vphantom {{{\rm{d}}\tilde r} {{\rm{d}}t}}} \right.
 \kern-\nulldelimiterspace} {{\rm{d}}t}} > 0\), yielding the following expression:
 \begin{eqnarray}\label{eq212}
{F_p} \approx 4\pi {\tilde r^2}{p_{ext}} & - &  \frac{{4{\rho _f}}}{{{r_{ch}}}}\left[ {\frac{{4\pi {L_{ch}}}}{{{r_{ch}}}} + \frac{{9\sqrt 2 }}{2}\left( {1 - {\mathop{\rm sgn}} \left( {\frac{{{\rm{d}}\tilde r}}{{{\rm{d}}t}}} \right)} \right)} \right]{\tilde r^4}\frac{{{{\rm{d}}^2}\tilde r}}{{{\rm{d}}{t^2}}}\nonumber\\ & - & \frac{{32{\rho _f}}}{{r_{ch}^2}}\left[ {\pi {L_{ch}} - \frac{{3{{\tilde r}^3}}}{{2r_{ch}^2}}\left( {1 - {\mathop{\rm sgn}} \left( {\frac{{{\rm{d}}\tilde r}}{{{\rm{d}}t}}} \right)} \right)} \right]{\tilde r^3}{\left( {\frac{{{\rm{d}}\tilde r}}{{{\rm{d}}t}}} \right)^2}.
\end{eqnarray}

\subsection{Numerical verification of the generalized force}\label{sec23}
To verify the expression of the generalized force, applied by the entrapped fluid on the extensional mode of the balloon, three different finite element schemes, allowing to examine the different terms in (\ref{eq212}), are simulated utilizing COMSOL Multiphysics. In all of these schemes, the flow is described according to the Navier-Stokes equations for incompressible fluids, assuming the processes are isothermal, where the density and dynamic viscosity of the fluid are taken as ${\rho _f} = 1000\left[ {{{kg} \mathord{\left/
 {\vphantom {{kg} {{m^3}}}} \right.
 \kern-\nulldelimiterspace} {{m^3}}}} \right]$,  ${\mu _f} = 8.94 \times {10^{ - 4}}\left[ {Pa \cdot s} \right]$.
 
 In the first scheme, the only effect being validated is the force applied directly by the external pressure, which is done by fixing the radii of the examined geometries, nulling all the forces in (\ref{eq212}), except for the first term. Namely, utilizing a numerical scheme where the external pressure varies according to a dictated profile, the generalized forces applied on different undeformed spherical geometries, which are connected to a channel, are examined. Several simulations carried out utilizing the abovementioned scheme show that the theoretical model captures almost flawlessly the effect of the external pressure. However, a visualization describing the comparison between the theoretical and the numerically obtained results is omitted for brevity.
 
\begin{figure}
\centering
\includegraphics[width=1\textwidth]{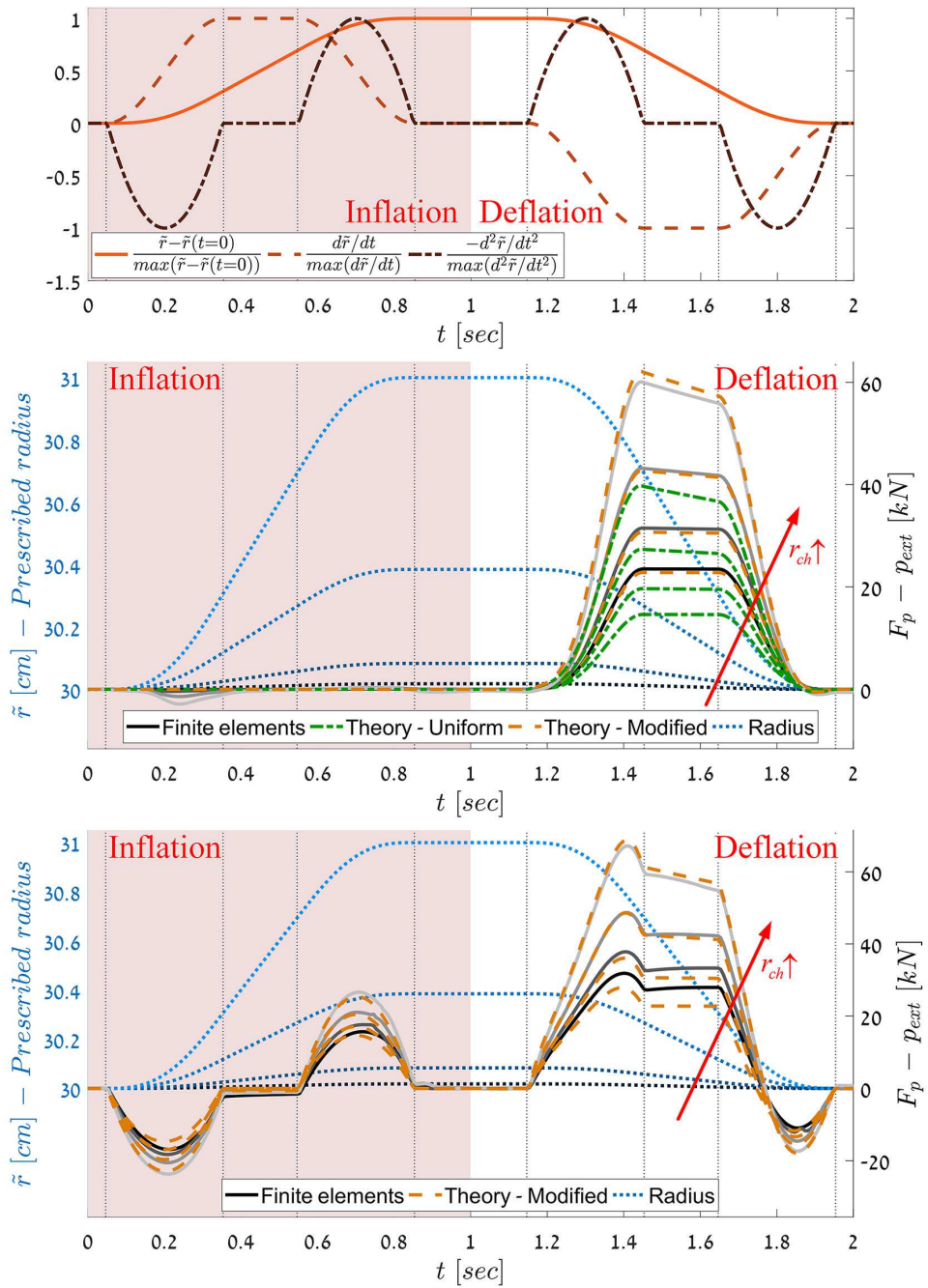}
\caption{Top: The normalized change of the prescribed radius (bright solid curve), radial velocity (dashed curve) and minus the radial acceleration (dark dash-dot curve) of the balloon's boundaries, corresponding to all simulations presented here. Middle and bottom: The non-static generalized forces acting on the different balloons in the case where the channel is absent and exists, respectively. The presented results are obtained from the finite element simulations (solid grey curves), the theoretical model (dash-dot green curves) and the modified model (dashed orange curves), when the instantaneous radii vary according to the dotted blue curves. In all cases, the brighter curves correspond to the geometries with the larger values of ${r_{ch}}$.}
\label{Figure3}
\end{figure}

 The next stage of the model verification is examining the terms, originating in the motion of the balloon. Here, since the two inertial forces and the two centripetal forces behave similarly, this stage is divided into two parts, allowing to distinct the terms originating in the flow inside the balloon from those, emanating from the flow inside the channel. Thus, in order to examine the remaining terms in (\ref{eq212}), two separate finite element schemes are utilized, simulating four different balloons with orifice radii of ${r_{ch}} = 0.5,\,1,\,2,\,3\,\left[ {cm} \right]$, whose boundaries vary according to dictated temporal profiles. In the first scheme, in order to isolate the forces, applied due to the flow inside the balloon, the channel is removed and the external pressure is nulled, where in the second scheme a channel with length of ${L_{ch}} = 50\left[ {cm} \right]$, is added in order to consider the contribution of the related terms. As shown in figure \ref{Figure3} top, presenting the normalized value and time derivatives corresponding to all temporal profiles, these consist both inflation and deflation, to examine the different behaviours exhibited by the fluid throughout the two regimes, see figure \ref{Figure2}. 
 
 Figure \ref{Figure3} middle, compares the generalized forces applied by the entrapped fluid on the four spherical balloons mentioned above, according to the theoretical model, and the numerical computations, while the channel is absent. This figure shows a substantial discrepancy between the numerical and theoretical results throughout deflation. Nevertheless, the deviations during inflation can be ignored, as the numerically calculated values are indeed negligible compared to those achieved throughout deflation. Comparison between the theoretical and the numerically calculated results, computed with and without a channel shows that the most substantial deviations are in the terms, related to the flow inside the balloon, with a strong emphasis on the centripetal force. The latter can be seen by observing the top and middle graphs in figure \ref{Figure3}, showing that the most significant deviations are related to the radial velocity.

 Thorough investigation shows that the major cause of the deviations throughout deflation is a discrepancy between the theoretical uniform flow out of the balloon, and the profile achieved numerically, which varies both spatially and temporally. Thus, in order to adjust the model to the more realistic case, exhibited by the numerical simulation, this profile should be considered. However, for sake of simplicity, the temporal variation of flow velocity profile, and its axial dependency along the channel, when exists, are assumed to be negligible, allowing to modify (\ref{eq21}) into the following form:
  \begin{equation}\label{eq213}
{\tilde v_z} = \hat f\left( \Upsilon  \right)\frac{{2 - \varepsilon }}{{\varepsilon \left( {1 - \varepsilon } \right)}}\frac{{{\rm{d}}\tilde r}}{{{\rm{d}}t}}.
\end{equation}
 Here, $\Upsilon$ is the radial coordinate of a cylindrical system, whose origin is located at the middle of the channel, normalized by ${r_{ch}}$. Over the imaginary continuation of the balloon, penetrating the channel, where the velocity boundary conditions are imposed, the relation of $\Upsilon $ with the non-inertial spherical system is given by \(\Upsilon  = {{\tilde r\sin \left( {\pi  - \theta } \right)} \mathord{\left/
 {\vphantom {{\tilde r\sin \left( {\pi  - \theta } \right)} {{r_{ch}}}}} \right.
 \kern-\nulldelimiterspace} {{r_{ch}}}}\). Furthermore, \(\hat f\left( \Upsilon  \right)\) denotes the axial flow velocity profile at the orifice, and all along the channel if the latter exists, which for sake of mass conservation, it is normalized such that \(\int\limits_0^1 {\Upsilon \hat f\left( \Upsilon  \right){\rm{d}}\Upsilon }  = \frac{1}{2}\). 
 
 Substituting (\ref{eq213}) into the unsteady Bernoulli equation, and comparing the stagnation pressure in both ends of the channel shows that the pressure field inside the channel varies along its radius. However, under the assumption that the channel is slender compared to the balloon's dimensions, this radial pressure variation is negligible, which is also supported by numerical results. In order to approximate the pressure inside the channel as radially uniform, the expression achieved from the unsteady Bernoulli equation is averaged along its radius, degenerating this expression into (\ref{eq23}). As a result, the first, second and fourth terms in (\ref{eq212}), obtained by imposing (\ref{eq23}) as the pressure at the middle of the balloon's inlet, are not affected by the model generalization. An additional corollary of the pressure averaging is that the model generalization affects only flow velocity boundary condition at the balloon's orifice, changing merely the values of the coefficients ${A_m}$. The modified coefficients, denoted ${\tilde A_m}$, are calculated in similar to their original form, while imposing the general boundary conditions, considering the non-uniform flow at the orifice, after development into a generalized Fourier series. This yields the following expression:
   \begin{equation}\label{eq214}
{\tilde A_m}\left( \varepsilon  \right) = \left\{ {\begin{array}{*{20}{c}}
{\frac{1}{{\varepsilon  - 1}}\left[ {\frac{{{\varepsilon ^2}\left( {3 - \varepsilon } \right)}}{2} + \frac{{3\left( {\varepsilon  - 2} \right)}}{\varepsilon }\int\limits_{\pi  - {{\sin }^{ - 1}}\sqrt {\varepsilon \left( {2 - \varepsilon } \right)} }^\pi  {\hat f\left( \Upsilon  \right)\sin \theta {{\cos }^2}\theta {\rm{d}}\theta } } \right]}&{m = 1}\\
{\frac{1}{{\varepsilon  - 1}}\left[ \begin{array}{l}
\frac{{{P_m}\left( {\varepsilon  - 1} \right) - \left( {\varepsilon  - 1} \right){P_{m - 1}}\left( {\varepsilon  - 1} \right)}}{{m - 1}} + \frac{{{P_m}\left( {\varepsilon  - 1} \right) - \left( {\varepsilon  - 1} \right){P_{m + 1}}\left( {\varepsilon  - 1} \right)}}{{m + 2}}\\
 + \frac{{\left( {\varepsilon-2 } \right)\left( {2m + 1} \right)}}{\varepsilon }\int\limits_{\pi  - {{\sin }^{ - 1}}\sqrt {\varepsilon \left( {2 - \varepsilon } \right)} }^\pi  {\hat f\left( \Upsilon  \right)\sin \theta \cos \theta {P_m}\left( {\cos \theta } \right){\rm{d}}\theta } 
\end{array} \right]}&{m = 2,3, \cdots }
\end{array}} \right.
\end{equation}
 Replacing the original values of ${A_m}$ by their modified values in (\ref{eqB}) allows repeating the numerical simplification process, modifying the functions ${g_3}$ and ${g_5}$, and therefore correcting the corresponding terms in (\ref{eq212}).
 
 By following the abovementioned algorithm, the modified forms of the functions ${g_3}$ and ${g_5}$ are calculated, where \(\hat f\left( \Upsilon  \right)\) is determined as a nominal profile achieved by averaging those, computed from all of the simulations, describing the system with and without a channel. In all cases, the profiles are acquired along the period where the balloon exhibits a constant, negative velocity. Moreover, for sake of simplicity the modified form of ${g_5}$ is calculated, considering only the first term inside the double series (see (\ref{eqB})) which is significantly larger compared to all other terms. The modified forms of ${g_3}$ and ${g_5}$ lead to corrections of both forces, originating in the flow inside the balloon, which after comparison to their original forms show relations with a weak dependency on $\varepsilon$. Thus, for simplicity, these relations are taken as constant values, such that in order to modify (\ref{eq212}), the original inertial force is multiplied by approximately 1.095, and the centripetal force which shows larger deviations, is multiplied by approximately 1.56. As a result, the corrected form of the generalized force, acting on the extensional mode of the balloon is:
\begin{eqnarray}\label{eq215}
{\tilde F_p} \approx 4\pi {\tilde r^2}{p_{ext}} & - & \frac{{4{\rho _f}}}{{{r_{ch}}}}\left[ {\frac{{4\pi {L_{ch}}}}{{{r_{ch}}}} + \frac{{223}}{{32}}\left( {1 - {\mathop{\rm sgn}} \left( {\frac{{{\rm{d}}\tilde r}}{{{\rm{d}}t}}} \right)} \right)} \right]{\tilde r^4}\frac{{{{\rm{d}}^2}\tilde r}}{{{\rm{d}}{t^2}}}\nonumber\\ & - & \frac{{32{\rho _f}}}{{r_{ch}^2}}\left[ {\pi {L_{ch}} - \frac{{75{{\tilde r}^3}}}{{32r_{ch}^2}}\left( {1 - {\mathop{\rm sgn}} \left( {\frac{{{\rm{d}}\tilde r}}{{{\rm{d}}t}}} \right)} \right)} \right]{\tilde r^3}{\left( {\frac{{{\rm{d}}\tilde r}}{{{\rm{d}}t}}} \right)^2},
\end{eqnarray}
where since it describes a more realistic situation, this formulation replaces its original form hereafter.

As seen from the middle and bottom graphs in figure \ref{Figure3}, the modified generalized force given by (\ref{eq215}) is highly correlated with its corresponding numerically calculated values. The latter implies that the deviations discussed above indeed arise from the discrepancy in the flow velocity boundary conditions at the balloon's orifice. Moreover, the good agreement throughout inflation, where all terms included in the generalized force emanates from the flow inside the channel, suggests that taking the pressure at the balloon's inlet as uniform is a valid assumption. Finally, it should be noted that in contrast to the well-ordered and irrotational flow velocity field presented in figure \ref{Figure2} for the case of deflation, in practice there is some residual vorticity originating in the inflation. Nevertheless, as seen in figure \ref{Figure3}, the latter does not cause noticeable deviations.

\subsection{Modelling the fully coupled system}
The final stage of the model derivation deals with formulating the fully coupled dynamics of the system. This is done by developing a variational model, describing the behaviour of the hyperelastic balloon, considering the influence of the entrapped fluid, discussed above. To simplify the model, and reduce the number of variables, the balloon's material is referred to as an incompressible solid, which is a valid assumption if for instance, the latter is a rubber balloon. An additional assumption, requiring that the thickness of the balloon is small, allows referring to it as a thin shell, meaning that its volume can be approximated by surface integration, multiplied by its thickness. Combination of the abovementioned assumptions gives rise a relation, connecting between the instantaneous radius and thickness $\tilde r,\,\tilde d$ and their unstarched values ${\tilde r_0},\,{\tilde d_0}$, whose leading order is given by \citep[][]{muller2004rubber}:
  \begin{equation}\label{eq216}
{\tilde r^2}\tilde d \approx \tilde r_0^2{\tilde d_0}.
\end{equation}
This relation allows eliminating the instantaneous thickness of the balloon, leaving the formulation of its fully coupled behaviour, dependent merely on its radius.

In order to formulate the dynamics of the balloon, its energies are derived under the assumptions mentioned above, starting with the kinetic energy related to its extensional motion. Considering only the leading order term, the kinetic energy is given by  \citep[][]{geradin2014mechanical}
  \begin{equation}\label{eq217}
{\cal T} \approx \frac{{{\rho _s}\tilde d}}{2}\mathop{{\int\!\!\!\!\!\int}\mkern-21mu \bigcirc}\limits_S 
 {\left( {\frac{{{\rm{d}}{{{\bf{\tilde r}}}^{\bf{T}}}}}{{{\rm{d}}t}} \cdot \frac{{{\rm{d}}{\bf{\tilde r}}}}{{{\rm{d}}t}}} \right){\rm{d}}S}  = 4\pi {\rho _s}\tilde r_0^2{\tilde d_0}{\left( {\frac{{{\rm{d}}\tilde r}}{{{\rm{d}}t}}} \right)^2} + O\left( \varepsilon  \right),
\end{equation}
where ${\rho _s}$ symbolizes the density of the solid, and \({{{\rm{d}}{\bf{\tilde r}}} \mathord{\left/
 {\vphantom {{{\rm{d}}{\bf{\tilde r}}} {{\rm{d}}t}}} \right.
 \kern-\nulldelimiterspace} {{\rm{d}}t}}\) denotes the velocity vector of each material point on the balloon, relative to a fixed reference point, which is calculated utilizing conservation of angular momentum along the axis of symmetry. 
 
 Next, for sake of formulating the potential energy of the balloon, its strain energy density function is derived, assuming it behaves as an incompressible, two parameter Mooney-Rivlin solid, where each material point undergoes equibiaxial tension, considering pure sphericity \citep[][]{muller2004rubber}:
   \begin{equation}\label{eq218}
W = {s_1}\left[ {2{{\left( {\frac{{\tilde r}}{{{{\tilde r}_0}}}} \right)}^2} + {{\left( {\frac{{{{\tilde r}_0}}}{{\tilde r}}} \right)}^4} - 3} \right] + {s_2}\left[ {{{\left( {\frac{{\tilde r}}{{{{\tilde r}_0}}}} \right)}^4} + 2{{\left( {\frac{{{{\tilde r}_0}}}{{\tilde r}}} \right)}^2} - 3} \right].
\end{equation}
Here, ${s_1},\,{s_2}$ are empirically determined material constants, and the stretches in the two different directions are given by \({{\tilde r} \mathord{\left/
 {\vphantom {{\tilde r} {{{\tilde r}_0}}}} \right.
 \kern-\nulldelimiterspace} {{{\tilde r}_0}}}\). Integrating (\ref{eq218}) over the volume of the balloon, and dropping high order terms yields the following expression, describing the leading order term of the system's potential energy \citep[][]{geradin2014mechanical}:
    \begin{equation}\label{eq219}
{\cal V} \approx \tilde d\mathop{{\int\!\!\!\!\!\int}\mkern-21mu \bigcirc}\limits_S 
 {W{\rm{d}}S}  = 4\pi \tilde r_0^2{\tilde d_0}W + O\left( \varepsilon  \right).
\end{equation}

The last expression needs to be formulated is the virtual work applied by the modified generalized force ${\tilde F_p}$ and the ambient pressure denoted ${p_a}$, whose leading order term is given as following:
    \begin{equation}\label{eq220}
\delta {\cal W} = \left[ {{{\tilde F}_p} - \mathop{{\int\!\!\!\!\!\int}\mkern-21mu \bigcirc}\limits_S 
 {{p_a}{\rm{d}}S} } \right]\delta \tilde r = \left[ {{{\tilde F}_p} - 4\pi {p_a}{{\tilde r}^2} + O\left( \varepsilon  \right)} \right]\delta \tilde r,
\end{equation}
where \(\delta \tilde r\) is a variation in \(\tilde r\).

The energies given by (\ref{eq217}) and (\ref{eq219}), supplemented by the virtual work (\ref{eq220}), describe the overall dynamics of the system in figure \ref{Figure1}, under the assumptions underlined above, while disregarding the internal damping of the balloon. Thus, applying the Hamilton's principle on (\ref{eq217}), (\ref{eq219}), (\ref{eq220}) yields the equation of motion, describing the balloon's extensional motion. In order to enable simplifying this equation by means of order of magnitude analysis, the following non-dimensional variables are introduced:
    \begin{equation}\label{eq221}
R = {{\tilde r} \mathord{\left/
 {\vphantom {{\tilde r} {{{\tilde r}_0}}}} \right.
 \kern-\nulldelimiterspace} {{{\tilde r}_0}}},\,\,\,\,\,\,P = {{\left( {{p_{ext}} - {p_a}} \right)} \mathord{\left/
 {\vphantom {{\left( {{p_{ext}} - {p_a}} \right)} {{p_{typ}}}}} \right.
 \kern-\nulldelimiterspace} {{p_{typ}}}},\,\,\,\,\,\,{S_{1,2}} = {{4{{\tilde d}_0}{s_{1,2}}} \mathord{\left/
 {\vphantom {{4{{\tilde d}_0}{s_{1,2}}} {{{\tilde r}_0}{p_{typ}}}}} \right.
 \kern-\nulldelimiterspace} {{{\tilde r}_0}{p_{typ}}}},\,\,\,\,\,\,T = {t \mathord{\left/
 {\vphantom {t {{t_{typ}}}}} \right.
 \kern-\nulldelimiterspace} {{t_{typ}}}},
\end{equation}
where ${p_{typ}}$ is a typical gauge pressure and ${t_{typ}}$ is a typical time scale. Substituting (\ref{eq221}) into the equation of motion yields the following non-dimensional form of this equation:
    \begin{equation}\label{eq222}
\begin{array}{l}
\left[ {\frac{{{\rho _s}}}{{{\rho _f}}}\frac{{{r_{ch}}{{\tilde d}_0}}}{{\tilde r_0^2}} + 2\frac{{{L_{ch}}}}{{{r_{ch}}}}{R^4} + \frac{{223}}{{64\pi }}{R^4}\left( {1 - {\mathop{\rm sgn}} \left( {\frac{{{\rm{d}}R}}{{{\rm{d}}T}}} \right)} \right)} \right]\frac{{{{\rm{d}}^2}R}}{{{\rm{d}}{T^2}}}\\
\,\,\,\,\,\,\,\,\,\,\,\,\,\,\,\,\,\,\,\,\,\,\,\,\,\,\,\,\,\,\, + \frac{{{r_{ch}}t_{typ}^2{p_{typ}}}}{{2{\rho _f}\tilde r_0^3}}\left[ {{S_1}\left( {R - {R^{ - 5}}} \right) + {S_2}\left( {{R^3} - {R^{ - 3}}} \right) - {R^2}P} \right]\\
\,\,\,\,\,\,\,\,\,\,\,\,\,\,\,\,\,\,\,\,\,\,\,\,\,\,\,\,\,\,\,\,\,\,\,\,\,\,\,\,\,\,\,\,\,\,\,\,\,\,\,\,\,\,\, =  - \left[ {4\frac{{{L_{ch}}}}{{{r_{ch}}}} - \frac{{75}}{{8\pi }}\frac{{\tilde r_0^3}}{{r_{ch}^3}}{R^3}\left( {1 - {\mathop{\rm sgn}} \left( {\frac{{{\rm{d}}R}}{{{\rm{d}}T}}} \right)} \right)} \right]{R^3}{\left( {\frac{{{\rm{d}}R}}{{{\rm{d}}T}}} \right)^2}.
\end{array}
\end{equation}
It should be noted that this expression is not written in its most concise form for sake of order of magnitude analysis, carried out in the upcoming section.

\subsubsection{Simplifying the fully coupled model}
In order to simplify the reduced-order model given by (\ref{eq222}), the terms, multiplying the different time derivatives of $R$ are compared, so small terms could be eliminated. This is done under the assumption that $R$ and its time derivatives are of order unity, due to appropriate normalization.

The first terms to be examined are those inside the first parenthesis in (\ref{eq222}), multiplying the second time derivative of $R$. As seen from (\ref{eq222}), the third of these terms, originating in the inertia of the flow inside the balloon, identically equals zero throughout inflation, whereas it is of order 1 during deflation. Further, the second term, originating in the inertia of the flow inside the channel, is of order \({{{L_{ch}}} \mathord{\left/
 {\vphantom {{{L_{ch}}} {{r_{ch}}}}} \right.
 \kern-\nulldelimiterspace} {{r_{ch}}}}\). And finally, the thickness of the balloon and the radius of the channel are significantly smaller, compared to the radius of the balloon. Thus, assuming the densities of the fluid and the solid are of the same order of magnitude, the first term, originating in the hyperelastic balloon's inertia, is of order \({{{r_{ch}}{{\tilde d}_0}} \mathord{\left/
 {\vphantom {{{r_{ch}}{{\tilde d}_0}} {\tilde r_0^2}}} \right.
 \kern-\nulldelimiterspace} {\tilde r_0^2}} \ll 1\). Accordingly, in the realistic case where \({{{L_{ch}}} \mathord{\left/
 {\vphantom {{{L_{ch}}} {{r_{ch}}}}} \right.
 \kern-\nulldelimiterspace} {{r_{ch}}}} \gg 1\), the second term is the prevailing force among the three discussed here, throughout both inflation and deflation. However, if \({{{L_{ch}}} \mathord{\left/
 {\vphantom {{{L_{ch}}} {{r_{ch}}}}} \right.
 \kern-\nulldelimiterspace} {{r_{ch}}}} \ll 1\), the third term is necessarily the governing force throughout deflation, where the prevailing force during inflation is determined based on the order of magnitude of \({{{L_{ch}}} \mathord{\left/
 {\vphantom {{{L_{ch}}} {{r_{ch}}}}} \right.
 \kern-\nulldelimiterspace} {{r_{ch}}}}\), compared to that of \({{{r_{ch}}{{\tilde d}_0}} \mathord{\left/
 {\vphantom {{{r_{ch}}{{\tilde d}_0}} {\tilde r_0^2}}} \right.
 \kern-\nulldelimiterspace} {\tilde r_0^2}}\).
 
 Next, the terms inside the third parenthesis, dependent on \({\left( {{{{\rm{d}}R} \mathord{\left/
 {\vphantom {{{\rm{d}}R} {{\rm{d}}T}}} \right.
 \kern-\nulldelimiterspace} {{\rm{d}}T}}} \right)^2}\) are examined. From the definition of $\varepsilon $ it is clear that \({{{{\tilde r}_0}} \mathord{\left/
 {\vphantom {{{{\tilde r}_0}} {{r_{ch}}}}} \right.
 \kern-\nulldelimiterspace} {{r_{ch}}}} \sim O\left( {{\varepsilon ^{ - {1 \mathord{\left/
 {\vphantom {1 2}} \right.
 \kern-\nulldelimiterspace} 2}}}} \right)\), meaning that the second term, originating in the centripetal force caused due to the flow inside the balloon is of order \({\varepsilon ^{ - {3 \mathord{\left/
 {\vphantom {3 2}} \right.
 \kern-\nulldelimiterspace} 2}}}\) throughout deflation, but it identically equals zero during inflation. Furthermore, the first term, representing the centripetal force induced due to the flow inside the channel, is of order \({{{L_{ch}}} \mathord{\left/
 {\vphantom {{{L_{ch}}} {{r_{ch}}}}} \right.
 \kern-\nulldelimiterspace} {{r_{ch}}}}\). Thus, when \({{{L_{ch}}} \mathord{\left/
 {\vphantom {{{L_{ch}}} {{r_{ch}}}}} \right.
 \kern-\nulldelimiterspace} {{r_{ch}}}} \gg {\varepsilon ^{ - {3 \mathord{\left/
 {\vphantom {3 2}} \right.
 \kern-\nulldelimiterspace} 2}}}\), the first term dominates throughout both inflation and deflation. However, if \({{{L_{ch}}} \mathord{\left/
 {\vphantom {{{L_{ch}}} {{r_{ch}}}}} \right.
 \kern-\nulldelimiterspace} {{r_{ch}}}} \ll {\varepsilon ^{ - {3 \mathord{\left/
 {\vphantom {3 2}} \right.
 \kern-\nulldelimiterspace} 2}}}\), the first term prevails throughout inflation, where it can be neglected compared to the second term during deflation.

The present paper deals with two limiting cases, corresponding to those examined in section \ref{sec23}. The first one is the simple situation where \({L_{ch}} \equiv 0\), meaning that the balloon is inflated through an orifice, in absence of a channel. According to the above discussion, the model in this case is degenerated into the following simplified form:
\begin{equation}\label{eq223}
\left\{ {\begin{array}{*{20}{c}}
{\frac{{{{\rm{d}}^2}R}}{{{\rm{d}}{T^2}}} + {C_1}\left[ {{S_1}\left( {{R^{ - 3}} - {R^{ - 9}}} \right) + {S_2}\left( {{R^{ - 1}} - {R^{ - 7}}} \right) - {R^{ - 2}}P} \right] = {C_2}{R^2}{{\left( {\frac{{{\rm{d}}R}}{{{\rm{d}}T}}} \right)}^2}}&{{{{\rm{d}}R} \mathord{\left/
 {\vphantom {{{\rm{d}}R} {{\rm{d}}T}}} \right.
 \kern-\nulldelimiterspace} {{\rm{d}}T}} < 0}\\
{\frac{{{{\rm{d}}^2}R}}{{{\rm{d}}{T^2}}} + {C_3}\left[ {{S_1}\left( {R - {R^{ - 5}}} \right) + {S_2}\left( {{R^3} - {R^{ - 3}}} \right) - {R^2}P} \right] = 0}&{{{{\rm{d}}R} \mathord{\left/
 {\vphantom {{{\rm{d}}R} {{\rm{d}}T}}} \right.
 \kern-\nulldelimiterspace} {{\rm{d}}T}} > 0,}
\end{array}} \right.
\end{equation}
where
 \refstepcounter{equation} \label{eq224}
$$
{C_1} = \frac{{16\pi {r_{ch}}t_{typ}^2{p_{typ}}}}{{223{\rho _f}\tilde r_0^3}},\,\,\,\,\,{C_2} = \frac{{600\tilde r_0^3}}{{223r_{ch}^3}},\,\,\,\,\,{C_3} = \frac{{t_{typ}^2{p_{typ}}}}{{2{\rho _s}{{\tilde r}_0}{{\tilde d}_0}}}.
\eqno{(\theequation{\mathit{a},\mathit{b},\mathit{c}})}
$$

The second, and more realistic case discussed in the scope of this paper is the situation where \(1 \ll {{{L_{ch}}} \mathord{\left/
 {\vphantom {{{L_{ch}}} {{r_{ch}}}}} \right.
 \kern-\nulldelimiterspace} {{r_{ch}}}} \ll {\varepsilon ^{ - {3 \mathord{\left/
 {\vphantom {3 2}} \right.
 \kern-\nulldelimiterspace} 2}}}\), reducing the model in (\ref{eq222}) to
\begin{equation}\label{eq225}
\frac{{{{\rm{d}}^2}R}}{{{\rm{d}}{T^2}}} + {\tilde C_1}\left[ {{S_1}\left( {{R^{ - 3}} - {R^{ - 9}}} \right) + {S_2}\left( {{R^{ - 1}} - {R^{ - 7}}} \right) - {R^{ - 2}}P} \right] = {\left( {\frac{{{\rm{d}}R}}{{{\rm{d}}T}}} \right)^2}\left\{ {\begin{array}{*{20}{c}}
{{{\tilde C}_2}{R^2}}&{{{{\rm{d}}R} \mathord{\left/
 {\vphantom {{{\rm{d}}R} {{\rm{d}}T < 0}}} \right.
 \kern-\nulldelimiterspace} {{\rm{d}}T < 0}}}\\
{ - 2{R^{ - 1}}}&{{{{\rm{d}}R} \mathord{\left/
 {\vphantom {{{\rm{d}}R} {{\rm{d}}T > 0,}}} \right.
 \kern-\nulldelimiterspace} {{\rm{d}}T > 0,}}}
\end{array}} \right.
\end{equation}
where
 \refstepcounter{equation} \label{eq226}
$$
{\tilde C_1} = \frac{{r_{ch}^2t_{typ}^2{p_{typ}}}}{{4{\rho _f}\tilde r_0^3{L_{ch}}}},\,\,\,\,\,\,{\tilde C_2} = \frac{{75\tilde r_0^3}}{{8\pi r_{ch}^2{L_{ch}}}}.
\eqno{(\theequation{\mathit{a},\mathit{b}})}
$$

\subsection{Numerical verification of the fully coupled model}\label{sec25}
\begin{figure}
\centering
\includegraphics[width=1\textwidth]{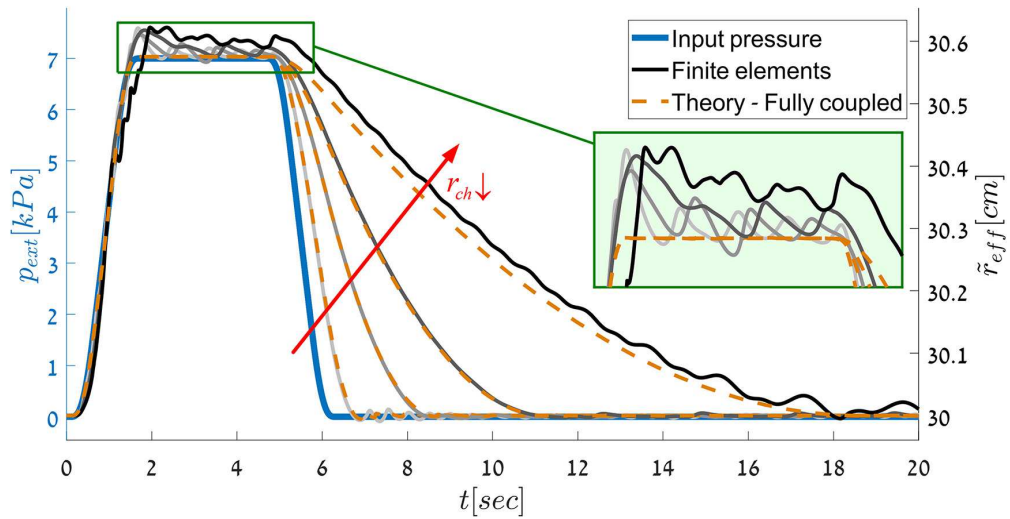}
\caption{The dynamic responses of the different balloons in the case where the channel is absent, according to the finite element simulations (solid grey curves), and the theoretical model (dashed orange curves), when the external pressure varies according to the solid blue curve. The darker numerically obtained curves correspond to the geometries with the smaller orifices.}
\label{Figure4}
\end{figure}

For sake of validating the fully coupled model derived above, its two degenerated variants, given by (\ref{eq223}) and (\ref{eq225}) are compared to finite element simulations carried out in COMSOL Multiphysics. In similar to section \ref{sec23}, in all of the simulations discussed here, the flow is described according to the Navier-Stokes equations for incompressible fluids, assuming the processes are isothermal, where the density and dynamic viscosity of the fluid are taken as ${\rho _f} = 1000\left[ {{{kg} \mathord{\left/
 {\vphantom {{kg} {{m^3}}}} \right.
 \kern-\nulldelimiterspace} {{m^3}}}} \right]$,  ${\mu _f} = 8.94 \times {10^{ - 4}}\left[ {Pa \cdot s} \right]$. Moreover, in correspondence to the theory, the balloon is modelled as a two parameter Moony-Rivlin solid whose material constants are \({s_1} = 1.5\left[ {MPa} \right]\), \({s_2} = 0.15\left[ {MPa} \right]\), where its density is taken as ${\rho _s} = 1000\left[ {{{kg} \mathord{\left/
 {\vphantom {{kg} {{m^3}}}} \right.
 \kern-\nulldelimiterspace} {{m^3}}}} \right]$. As in section \ref{sec23}, and in order to validate both degenerated models, two distinct numerical schemes are executed; In the first one, the balloon is inflated through a small orifice, in absence of a channel, whereas the second scheme is supplemented by a channel whose length is ${L_{ch}} = 50\left[ {cm} \right]$. In both schemes, the unstretched radius and thickness of the balloon are taken as ${\tilde r_0} = 30\left[ {cm} \right]$, ${\tilde d_0} = 0.3\left[ {cm} \right]$, and the maximal radius of the channel is set to be ${r_{ch}} = 3\left[ {cm} \right]$, which together with all other parameters discussed above, meet the assumptions, degenerating (\ref{eq222}) into (\ref{eq223}) and (\ref{eq225}). As seen in figures \ref{Figure4} and \ref{Figure6}, in each simulation, the external pressure is increased and decreased according to a similar profile along the lower stable monotonic branch of the solid's stress-strain relation, see figure \ref{Figure7}. This allows examining the assumptions on which the reduced order model is based, throughout both inflation and deflation, with an emphasis on the assumption that the balloon stays approximately spherical throughout its motion. 

\begin{figure}
\centering
\includegraphics[width=0.49\textwidth]{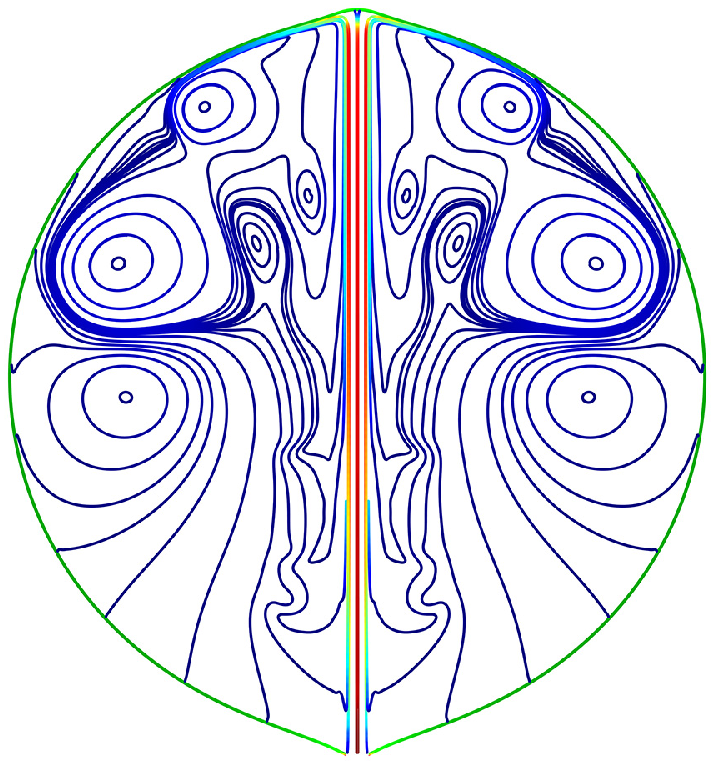}
\includegraphics[width=0.49\textwidth]{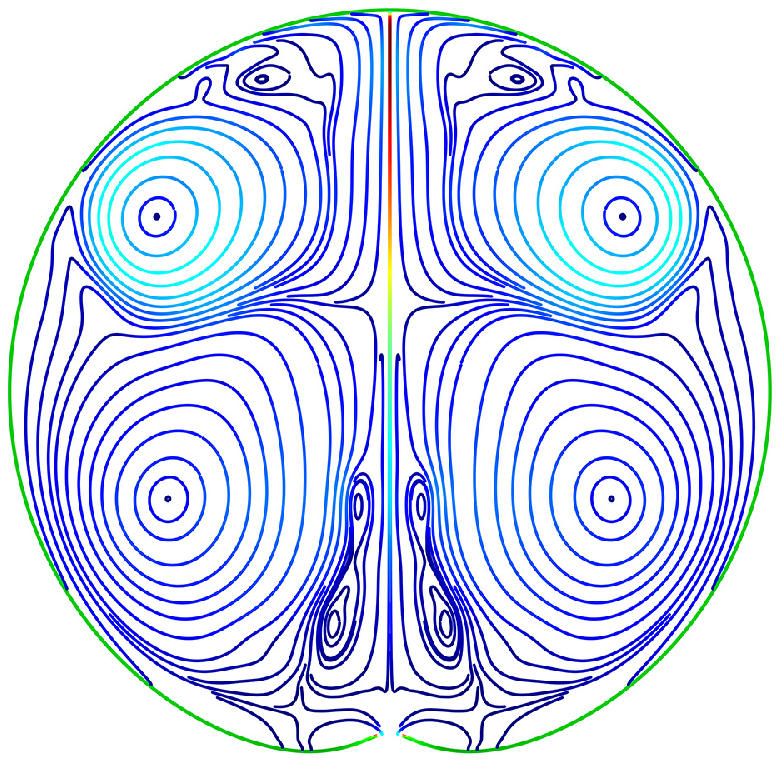}
\caption{Typical deformation of a balloon alongside the streamlines, describing the flow velocity field of the entrapped fluid, throughout a rapid inflation, in two extreme instances. The latter shows the non-spherical stretching of the balloon, caused thanks to the impinging jet (left), and the resultant pear-shaped deformation, achieved when the balloon is released from this stretching, after the jet vanishes (right). The red regions of the streamlines correspond to the highest flow velocities, whereas the blue regions represent the lowest flow velocities.}
\label{Figure5}
\end{figure}

\begin{figure}
\centering
\includegraphics[width=1\textwidth]{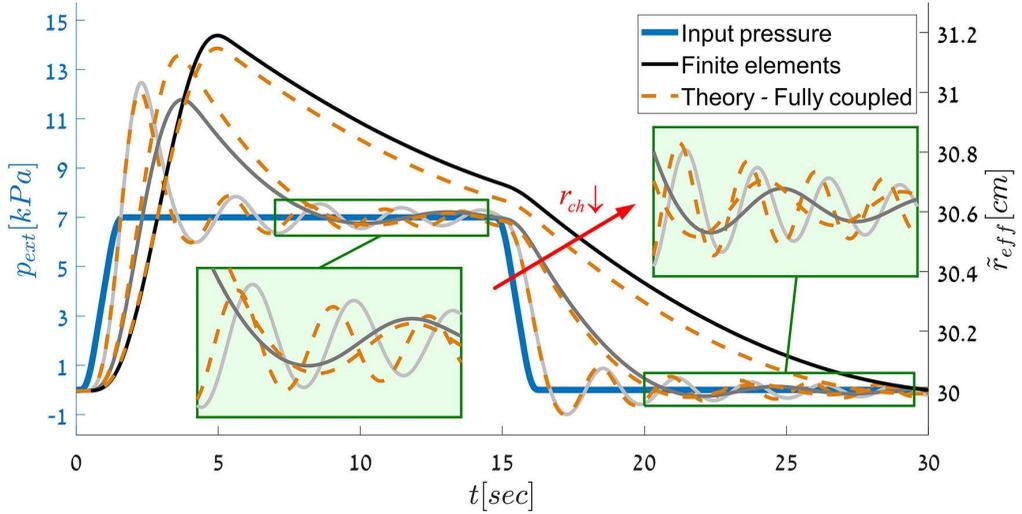}
\caption{The dynamic responses of the different balloons in the case where the channel exists, according to the finite element simulations (solid grey curves), and the theoretical model (dashed orange curves), when the external pressure varies according to the solid blue curve. The darker numerically obtained curves correspond to the geometries with the narrower channels.}
\label{Figure6}
\end{figure}

Figure \ref{Figure4} compares the theoretical and the numerically obtained dynamic responses of four different balloons whose orifice radii are ${r_{ch}} = 1,\,\,1.5,\,\,2,\,\,3\left[ {cm} \right]$, which are subjected to the same time varying external pressure, applied directly at their inlets. Here, since in the finite element simulations, the balloons are not restricted to be spherical, the numerical results are given in terms of the effective radius of the balloon, defined by \({\tilde r_{eff}} \buildrel \Delta \over = {{\sqrt {{S \mathord{\left/
 {\vphantom {S \pi }} \right.
 \kern-\nulldelimiterspace} \pi } + r_{ch}^2} } \mathord{\left/
 {\vphantom {{\sqrt {{S \mathord{\left/
 {\vphantom {S \pi }} \right.
 \kern-\nulldelimiterspace} \pi } + r_{ch}^2} } 2}} \right.
 \kern-\nulldelimiterspace} 2}\). The latter is formulated based on the theoretical relation between the measured inner surface area of the balloon denoted $S$, and the balloon's radius, after simplification utilizing the second order Taylor approximation of $\varepsilon$, around ${{{r_{ch}}} \mathord{\left/
 {\vphantom {{{r_{ch}}} {\tilde r}}} \right.
 \kern-\nulldelimiterspace} {\tilde r}} = 0$. Figure \ref{Figure4} shows a good correlation between the numerical and the theoretical results throughout both inflation and deflation. Here, as expected, since the the centripetal force acting against the balloon's motion throughout deflation grows when the radius of the orifice becomes smaller, the balloons with the smaller orifices experience slower deflation. Moreover, since the only substantial force applied by the fluid throughout inflation is the one emanating from the external pressure, the balloons inflate in the same rate as the pressure increment, as predicted by the theory. However, there are some discrepancies shown in figure \ref{Figure4}. These inconsistencies originate in non-spherical modes that are not considered in the theoretical model, which also distort the pressure field, compared to its theoretical form. An example for this phenomenon is demonstrated in figure \ref{Figure5} and \citep[][]{tezduyar2007modelling}, showing that in practice, the internal jet formed throughout inflation, stretch the balloon non-spherically, while after the jet vanishes, the balloon does not immediately return to its original shape. Instead, it oscillates between this stretched shape and a pear shape, until convergence to a steady sphere. This phenomenon becomes stronger at high inflation rates and in balloons with smaller orifice sizes, since in these cases the forces applied by the impinging jet become stronger and more focused.

Next, figure \ref{Figure6} examines the agreement between the theoretical and the simulated dynamic responses of three balloons, connected to channels with the same length which equals to ${L_{ch}} = 50\left[ {cm} \right]$, but with different radii, given by ${r_{ch}} = 1,\,\,1.5,\,\,3\left[ {cm} \right]$. Here, the external pressure applied at the inlet of the channel, varies at a similar profile to the one utilized in the case where the channel is absent, but with a longer delay between inflation and deflation. Comparison between the analyses shown in figures \ref{Figure4} and \ref{Figure6} shows an inverse relation between the excitation of the non-spherical modes, and the length of the channel. The latter is since the forces, resisting the balloon's inflation thanks to the flow inside the channel increase with ${L_{ch}}$. Thus, longer channels lead to slower inflation and thus to lower deviations from the theory due to a weaker excitation of non-spherical modes. Clearly, reducing the radius of the channel also decrease the inflation rate. Thus, although narrower channels lead to more focused jets, they also make them weaker. Accordingly, the agreement between the numerical and the theoretical results shown in figure \ref{Figure6} does not have a monotonic relation with the radius of the channel. Nevertheless, the results given in both figures show that the theoretical model gives a good prediction of the system's dynamics. However, it should still be used cautiously, especially at high inflation rates, and when the channel, connecting the balloon to the pressure inlet, is short and narrow. 

\section{Case Studies}
The upcoming sections deal with investigating the different variants of the fully coupled model (\ref{eq222}), providing insights and shedding light on their static and dynamic behaviour. As discussed in the previous section, since the investigated systems are derived under simplifying assumptions, in practice, the results presented below are expected to be distorted by unmodelled effects.

\subsection{Static behaviour}
The first analysis discussed here deals with the static behaviour of the reduced order model (\ref{eq222}) and all of its simplified variants, describing the unforced system after convergence to a stable equilibrium radius. To investigate this static behaviour, the time derivatives of $R$ are nulled, leading to the following equation, connecting between the equilibrium radii ${R_{eq}}$, and the constant pressure at the input and inside the balloon, given by \({P_0}\): 
  \begin{equation}\label{eq31}
{P_0} = {S_1}\left( {R_{eq}^{ - 1} - R_{eq}^{ - 7}} \right) + {S_2}\left( {{R_{eq}} - R_{eq}^{ - 5}} \right).
\end{equation}

\begin{figure}
\centering
\includegraphics[width=0.49\textwidth]{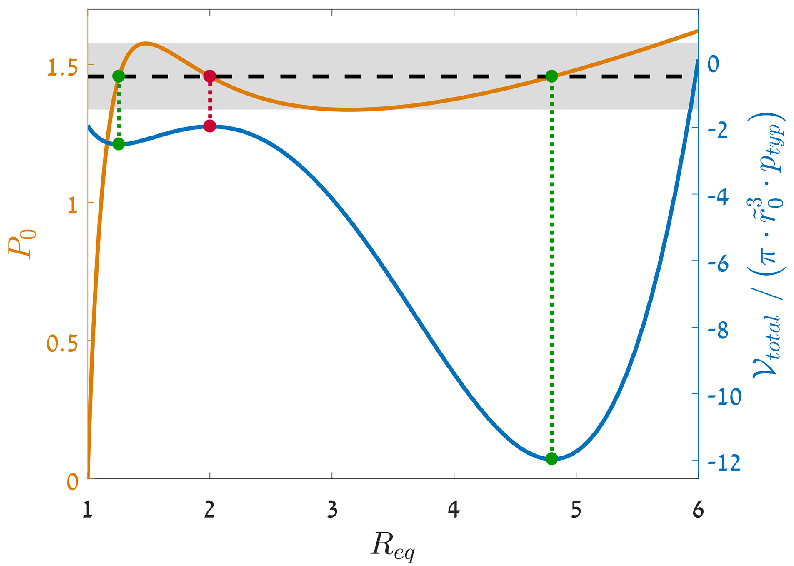}
\includegraphics[width=0.49\textwidth]{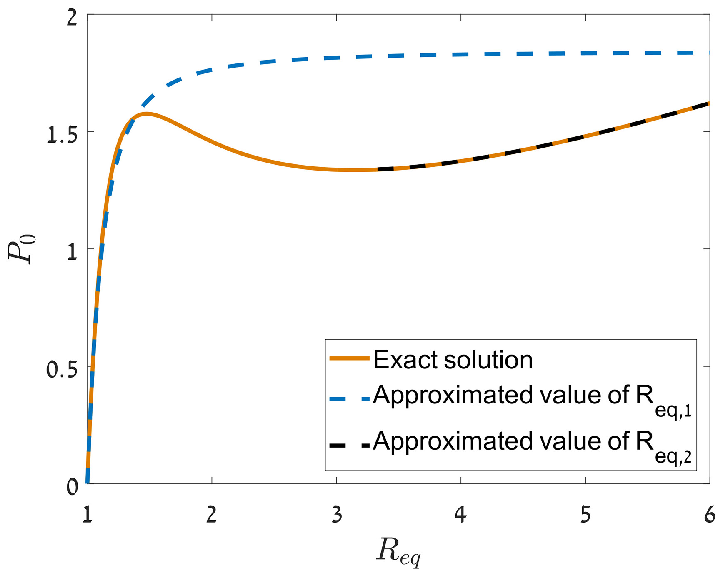}
\caption{Left: Solid orange curve -- A typical relation between the static pressure and the equilibrium radius of a spherical balloon; Solid blue curve -- The normalized potential energy function, corresponding to the constant pressure, illustrated by the dashed black line; Green and red dots -- The stable and unstable equilibrium radii.  Right: Blue and black dashed curves -- The approximated equilibrium radii, according to the asymptotic analyses; Solid orange curve -- The exact relation between the static pressure and the equilibrium radius.}
\label{Figure7}
\end{figure}

Figure \ref{Figure7} left presents a typical, non-monotonic relation between the constant pressure inside a balloon, and the corresponding equilibrium radii according to (\ref{eq31}), illustrated utilizing the same parameters as in section \ref{sec25}, where ${{{S_1}} \mathord{\left/
 {\vphantom {{{S_1}} {{S_2}}}} \right.
 \kern-\nulldelimiterspace} {{S_2}}} = 10$. This figure shows a bifurcation, which occurs when the pressure enters or exits the range between the local extrema, illustrated in grey. In this range, every constant pressure leads to two stable equilibrium radii, and one unstable radius, while above or below it there is only one stable equilibrium. The abovementioned behaviour can be demonstrated by the overall potential energy of the system, achieved by combining the potential energy of the balloon (\ref{eq219}), and those related to the entrapped and the ambient fluids, yielding the following expression:
   \begin{equation}\label{eq32}
{{\cal V}_{total}} \approx \pi \tilde r_0^3{p_{typ}}\left[ {{S_1}\left( {2R_{eq}^2 + R_{eq}^{ - 4} - 3} \right) + {S_2}\left( {R_{eq}^4 + 2R_{eq}^{ - 2} - 3} \right) - \frac{4}{3}R_{eq}^3{P_0}} \right].
\end{equation}
Indeed, Based on (\ref{eq32}), figure \ref{Figure7} left shows a double-well potential energy function, indicating on two stable equilibria, obtained when the constant pressure is in the grey region.

It can be seen that the relation between ${P_0}$ and ${R_{eq}}$ given by (\ref{eq31}), is a transcendental equation, meaning that closed form solutions of the equilibrium radii cannot be achieved analytically. However, utilizing some insights from figure \ref{Figure7} can aid in finding approximations for the stable equilibria. Namely, figure \ref{Figure7} suggests that when the pressure does not exceed the local maximum, there is a stable equilibrium radius around ${R_{eq}} = 1$. In order to find an approximation for this smaller stable equilibrium radius, a small parameter $\delta  \ll 1$ is defined, and is utilized to describe the latter as a second order asymptotic series given by ${R_{eq,1}} = 1 + \delta  + {\delta ^2} + O\left( {{\delta ^3}} \right)$. Substituting this series into (\ref{eq31}) after multiplication by $R_{eq}^7$, and dropping all terms of order ${\delta ^3}$ and higher, yields a second order equation in terms of $\delta $, whose solutions are given by
 \begin{equation}\label{eq33}
{\delta _{1,2}} = \frac{{6\left( {{S_1} + {S_2}} \right) - 7{P_0} \pm \sqrt {36{{\left( {{S_1} + {S_2}} \right)}^2} + 48{P_0}{S_2} - 63P_0^2} }}{{56{P_0} - 42{S_1} - 66{S_2}}}.
\end{equation}
Here, since $\delta $ is necessarily non-negative, the corresponding solution, having a negative sign in front of the square root of the discriminant, is taken. Substituting the chosen solution into the abovementioned series, yields a second order approximation of the smaller stable equilibrium radius.

Next, according to figure \ref{Figure7}, the second stable equilibrium radius, which exists when the pressure is higher than the local minimum, is considered significantly larger than unity. Thus, in order to formulate an approximation for this larger stable equilibrium radius, ${R_{eq,2}} = {\delta ^{ - 1}} \gg 1$ is substituted into (\ref{eq31}), yielding the following relation:
 \begin{equation}\label{eq34}
{P_0} = {S_1}\delta  + {S_2}{\delta ^{ - 1}} + O\left( {{\delta ^5}} \right),
\end{equation}
which after division by $\delta $ and returning to the original variables, yields a second order algebraic equation in terms of ${R_{eq,2}}$. The solution of this equation is given by
 \begin{equation}\label{eq35}
{\left( {{R_{eq,2}}} \right)_{1,2}} = \frac{{{P_0} \pm \sqrt {P_0^2 - 4{S_1}{S_2}} }}{{2{S_2}}},
\end{equation}
where in order to strengthen the assumption that \({R_{eq,2}} \gg 1\), the larger solution, having a positive sign in front of the square root of the discriminant, is chosen.

Figure \ref{Figure7} right compares the approximated solutions of the stable equilibria, with the exact relation between pressure and the equilibrium radii, given by (\ref{eq31}), where the different approximations are considered only when they are real-valued. This figure shows that the approximations of both stable equilibrium radii provide very good estimations in the regions where these equilibria  exist. Moreover, outside the validity region of ${R_{eq,2}}$, its approximation attains complex values, which can be utilized to assess the region where this solution exists. However, outside the validity region of ${R_{eq,1}}$, its approximation stays real-valued, what necessitates to estimate this region using a different method, such as estimating the local maximum.

\subsection{Order of magnitude analysis}
The current section serves as a preparation for the local asymptotic analyses discussed below, which require defining a small parameter. For this sake, since the only two systems investigated here are those given by (\ref{eq223}) and (\ref{eq225}), the orders of magnitude of their different constant coefficients are evaluated, around their stable equilibria.

\subsubsection{No channel}
The first degenerated system analysed here is the one given by (\ref{eq223}), where the channel is absent. In order to evaluate the different coefficients of this system, first the typical time constant ${t_{typ}}$, corresponding to each piecewise equation is estimated as the inverse value of the natural frequency, around a stable equilibrium. Namely, the typical time constants are calculated by inverting the values, achieved by developing the dimensional form of (\ref{eq223}) into first order Taylor series around a general stable equilibrium point, and taking the square root of the coefficients of $\Delta \tilde r = \tilde r - {\tilde r_{eq}}$. The latter is done while referring to the gauge pressure as a constant, nominal value ${P_0}$, while utilizing (\ref{eq31}). The time constants achieved by these means convert the coefficients ${C_1}$ and ${C_3}$ into the following forms
 \begin{subequations}\label{eq36}
 \begin{equation}
{C_1} = {\left[ {{S_1}\left( { - R_{eq}^{ - 4} + 7R_{eq}^{ - 10}} \right) + {S_2}\left( {R_{eq}^{ - 2} + 5R_{eq}^{ - 8}} \right)} \right]^{ - 1}},
\end{equation}
 \begin{equation}
{C_3} = {\left[ {{S_1}\left( { - 1 + 7R_{eq}^{ - 6}} \right) + {S_2}\left( {R_{eq}^2 + 5R_{eq}^{ - 4}} \right)} \right]^{ - 1}},
\end{equation}
\end{subequations}
which are of order 1 since the non-dimensional variables are considered of order unity, due to appropriate normalization. However, recalling that  \({{{{\tilde r}_0}} \mathord{\left/
 {\vphantom {{{{\tilde r}_0}} {{r_{ch}}}}} \right.
 \kern-\nulldelimiterspace} {{r_{ch}}}} \sim O\left( {{\varepsilon ^{ - {1 \mathord{\left/
 {\vphantom {1 2}} \right.
 \kern-\nulldelimiterspace} 2}}}} \right)\), the order of magnitude of ${C_2}$ is \({\varepsilon ^{ - {3 \mathord{\left/
 {\vphantom {3 2}} \right.
 \kern-\nulldelimiterspace} 2}}}\). Thus, the small parameter used when dealing with the system in (\ref{eq223}), is defined by $\delta  \buildrel \Delta \over = C_2^{ - 1} \ll 1$.
 
 \subsubsection{Long channel}
 A similar procedure for finding the typical time constant of the system in (\ref{eq225}), around a stable equilibrium radius yields a value, converting ${\tilde C_1}$ into the following form:
 \begin{equation}\label{eq37}
{\tilde C_1} = {\left[ {{S_1}\left( { - R_{eq}^{ - 4} + 7R_{eq}^{ - 10}} \right) + {S_2}\left( {R_{eq}^{ - 2} + 5R_{eq}^{ - 8}} \right)} \right]^{ - 1}},\,
\end{equation}
 which is again, of order unity, assuming appropriate normalization of the non-dimensional variables. Nevertheless, since in this case \(1 \ll {{{L_{ch}}} \mathord{\left/
 {\vphantom {{{L_{ch}}} {{r_{ch}}}}} \right.
 \kern-\nulldelimiterspace} {{r_{ch}}}} \ll {\varepsilon ^{ - {3 \mathord{\left/
 {\vphantom {3 2}} \right.
 \kern-\nulldelimiterspace} 2}}}\), and as abovementioned \({{{{\tilde r}_0}} \mathord{\left/
 {\vphantom {{{{\tilde r}_0}} {{r_{ch}}}}} \right.
 \kern-\nulldelimiterspace} {{r_{ch}}}} \sim O\left( {{\varepsilon ^{ - {1 \mathord{\left/
 {\vphantom {1 2}} \right.
 \kern-\nulldelimiterspace} 2}}}} \right)\), then \({\tilde C_2} \gg 1\). Thus, when analysing the system given by (\ref{eq225}), the small parameter utilized is defined by $\tilde \delta  \buildrel \Delta \over = \tilde C_2^{ - 1} \ll 1$.
 
\subsection{Free motion around a stable equilibrium radius}\label{sec33}
The current section deals with the free motion of the degenerated systems (\ref{eq223}) and (\ref{eq225}), in close proximity to a stable equilibrium radius, assuming they start from rest, while the pressure at the inlet is constant and equals to \({P_0}\).
 
\subsubsection{No channel}\label{sec331}
In similar to the previous section, the first analysis deals with the system given by (\ref{eq223}), where the channel is absent. The investigation of this system begins with a perturbation analysis of the equation, describing its behaviour throughout deflation, providing an approximated solution of its unforced dynamics near a stable equilibrium radius. The latter is followed by an analysis, leading to the corresponding approximated behaviour during inflation. Finally, both approximations are combined, yielding a uniform solution, describing the overall unforced dynamics of (\ref{eq223}), near a stable equilibrium.
  
Formulating the equation, describing the deflation dynamics of (\ref{eq223}) in terms of $\delta $, shows that the highest time derivative is not included in the leading order. Thus, for sake of finding an approximated solution for this equation, matched asymptotic expansion is utilized \citep[][]{holmes2012introduction,nayfeh2008perturbation}. Namely, the analysis of the deflation equation is carried out over two stages; the first one describes a relatively slow transient response, which does not include the initial moment, whereas the second stage deals with the fast transient, describing the initial response. Combination of these expressions leads to a uniform approximated solution of the deflation dynamics.
  
As abovementioned, the first part of the analysis of the deflation equation, deals with the slower transient, also referred to as the outer solution of the equation. To find an approximation for this response, a regular perturbation analysis is applied, utilizing the following asymptotic series:
\begin{equation}\label{eq38}
{R_{outer}}\left( T \right) = {R_0}\left( T \right) + {\delta ^{{1 \mathord{\left/
 {\vphantom {1 2}} \right.
 \kern-\nulldelimiterspace} 2}}}{R_1}\left( T \right) + \delta {R_2}\left( T \right) + {\delta ^{{3 \mathord{\left/
 {\vphantom {3 2}} \right.
 \kern-\nulldelimiterspace} 2}}}{R_3}\left( T \right) + O\left( {{\delta ^2}} \right).
\end{equation}
Thus, (\ref{eq38}) is substituted into the investigated equation, yielding an expression, which consists terms of different orders of $\delta $. This allows separating the abovementioned formulation into multiple equations, which are consecutively solved starting from the leading order equation, consisting all the terms of order unity, until the third order equation, which include the terms of order ${\delta ^2}$. The different equations lead to the formulations of ${R_0} \div {R_3}$ whose substitution into (\ref{eq38}) provides the following outer solution:
\begin{equation} \label{eq39}
\begin{array}{l}
{R_{outer}}\left( T \right) = {B_0} + {\delta ^{{1 \mathord{\left/
 {\vphantom {1 2}} \right.
 \kern-\nulldelimiterspace} 2}}}\left[ {{B_1} - \sqrt {{\beta _1}} T} \right] + \delta \left[ {{B_2} - {B_1}{\beta _2}T + \frac{{\sqrt {{\beta _1}} {\beta _2}}}{2}{T^2}} \right]\\
 + {\delta ^{{3 \mathord{\left/
 {\vphantom {3 2}} \right.
 \kern-\nulldelimiterspace} 2}}}\left[ {{B_3} - \frac{{\sqrt {{\beta _1}} {\beta _2}\left( {B_0^{ - 2} + 2{B_2}} \right) + B_1^2\left( {{\beta _3} - \beta _2^2} \right)}}{{2\sqrt {{\beta _1}} }}T + \frac{{{B_1}{\beta _3}}}{2}{T^2} - \frac{{\sqrt {{\beta _1}} {\beta _3}}}{6}{T^3}} \right] + O\left( {{\delta ^2}} \right),
\end{array}
\end{equation}
where ${B_0} \div {B_3}$ are unknown constant values which are to be determined in the matching stage, and ${\beta _1} \div {\beta _3}$ are constant expressions, given by:
\begin{subequations}\label{eq310}
\begin{equation}
{\beta _1} = {C_1}\left[ {{S_1}\left( {B_0^{ - 5} - B_0^{ - 11}} \right) + {S_2}\left( {B_0^{ - 3} - B_0^{ - 9}} \right) - B_0^{ - 4}{P_0}} \right],
\end{equation}
\begin{equation}
{\beta _2} = {C_1}\frac{{{S_1}\left( { - 5B_0^{ - 6} + 11B_0^{ - 12}} \right) + 3{S_2}\left( { - B_0^{ - 4} + 3B_0^{ - 10}} \right) + 4B_0^{ - 5}{P_0}}}{{2\sqrt {{\beta _1}} }},
\end{equation}
\begin{equation}
{\beta _3} = {C_1}\left[ {3{S_1}\left( {5B_0^{ - 7} - 22B_0^{ - 13}} \right) + 3{S_2}\left( {2B_0^{ - 5} - 15B_0^{ - 11}} \right) - 10B_0^{ - 6}{P_0}} \right].
\end{equation}
\end{subequations}

The second stage of the derivation deals with the inner solution, describing the initial moment of the system's deflation dynamics. For this sake, a faster time scale defined by $\tau  = {\delta ^{ - {1 \mathord{\left/
 {\vphantom {1 2}} \right.
 \kern-\nulldelimiterspace} 2}}}T$, and the asymptotic series

\begin{equation} \label{eq311}
{R_{inner}}\left( \tau  \right) = {R_0}\left( \tau  \right) + \delta {R_1}\left( \tau  \right) + {\delta ^2}{R_2}\left( \tau  \right) + O\left( {{\delta ^3}} \right),
\end{equation}
are substituted into the equation under investigation, yielding an expression, consisting of several orders of $\delta $. In similar to the derivation of the outer solution, the equations corresponding to the different orders of $\delta $ are consecutively solved, together with the initial conditions \({R_{inner}}\left( {\tau  = 0} \right) = {R_i},\,\,\,\,\,\,{\left. {{{{\rm{d}}{R_{inner}}} \mathord{\left/
 {\vphantom {{{\rm{d}}{R_{inner}}} {{\rm{d}}\tau }}} \right.
 \kern-\nulldelimiterspace} {{\rm{d}}\tau }}} \right|_{\tau  = 0}} = 0\), yielding the closed form formulations of ${R_0} \div {R_2}$, under the assumption that the system starts from rest. Substituting these expressions into (\ref{eq311}) provides the following inner solution:

\begin{equation} \label{eq312}
\begin{array}{l}
{R_{inner}}\left( \tau  \right) = {R_i} - \delta \frac{{\ln \left( {\cosh \left( {{R_i}\sqrt {{\gamma _1}} \tau } \right)} \right)}}{{R_i^2}}\\
\,\,\,\,\,\,\,\,\,\,\,\,\,\,\,\,\,\,\,\,\,\,\,\,\,\,\,\,\,\,\, - {\delta ^2}\frac{{{R_i}\sqrt {{\gamma _1}} \tau \tanh \left( {{R_i}\sqrt {{\gamma _1}} \tau } \right) + \ln \left( {\cosh \left( {{R_i}\sqrt {{\gamma _1}} \tau } \right)} \right)\left[ {\ln \left( {\cosh \left( {{R_i}\sqrt {{\gamma _1}} \tau } \right)} \right) - 2} \right]}}{{R_i^5}}\\
\,\,\,\,\,\,\,\,\,\,\,\,\,\,\,\,\,\,\,\,\,\,\,\,\,\,\,\,\,\, - {\delta ^2}\frac{{{\gamma _2}}}{{4R_i^5{\gamma _1}}}\left\{ \begin{array}{l}
\tanh \left( {{R_i}\sqrt {{\gamma _1}} \tau } \right)\left[ \begin{array}{l}
L{i_2}\left( { - {{\rm{e}}^{ - 2{R_i}\sqrt {{\gamma _1}} \tau }}} \right)\\
 + \frac{{{\pi ^2}}}{{12}} + R_i^2{\gamma _1}{\tau ^2}\\
 + {R_i}\sqrt {{\gamma _1}} \tau \left( {1 - 2\ln 2} \right)
\end{array} \right]\\
 + \frac{{{\pi ^2}}}{6} + L{i_2}\left( { - {{\rm{e}}^{ - 2{R_i}\sqrt {{\gamma _1}} \tau }}} \right) + L{i_2}\left( { - {{\rm{e}}^{2{R_i}\sqrt {{\gamma _1}} \tau }}} \right)\\
 + 2R_i^2{\gamma _1}{\tau ^2} - 2\ln \left( {\cosh \left( {{R_i}\sqrt {{\gamma _1}} \tau } \right)} \right)
\end{array} \right\} + O\left( {{\delta ^3}} \right),
\end{array}
\end{equation}
where $L{i_n}\left(  \bullet  \right)$ is a polylogarithm of order $n$, and ${\gamma _1},\,\,{\gamma _2}$ are constant expressions, given by
\begin{subequations}\label{eq313}
\begin{equation}
{\gamma _1} = {C_1}\left[ {{S_1}\left( {R_i^{ - 3} - R_i^{ - 9}} \right) + {S_2}\left( {R_i^{ - 1} - R_i^{ - 7}} \right) - R_i^{ - 2}{P_0}} \right],
\end{equation}
\begin{equation}
{\gamma _2} = {C_1}\left[ {{S_1}\left( {R_i^{ - 3} - 7R_i^{ - 9}} \right) - {S_2}\left( {R_i^{ - 1} + 5R_i^{ - 7}} \right)} \right].
\end{equation}
\end{subequations}

The third stage of the deflation dynamics formulation is a matching procedure, whose purpose is to derive a uniform expression based on both inner and outer solutions. For this, since the inner solution is unbounded as $\tau  \to \infty $, ${R_{inner}}$ and ${R_{outer}}$ are matched at this limit, where the inner solution degenerates into a relatively simple asymptotic approximation, given in terms of $T$ by:
\begin{equation}\label{eq314}
\begin{array}{l}
{R_{inner}}\left( {\tau  \gg 1} \right) = {R_i} - {\delta ^{{1 \mathord{\left/
 {\vphantom {1 2}} \right.
 \kern-\nulldelimiterspace} 2}}}\frac{{\sqrt {{\gamma _1}} }}{{{R_i}}}T + \delta \left[ {\frac{{\ln 2}}{{R_i^2}} - \frac{{4{\gamma _1} + {\gamma _2}}}{{4R_i^3}}{T^2}} \right] + {\delta ^{{3 \mathord{\left/
 {\vphantom {3 2}} \right.
 \kern-\nulldelimiterspace} 2}}}\frac{{\left( {4{\gamma _1} + {\gamma _2}} \right)\left( {1 + 2\ln 2} \right)}}{{4R_i^4\sqrt {{\gamma _1}} }}T\\
\,\,\,\,\,\,\,\,\,\,\,\,\,\,\,\,\,\,\,\,\,\,\,\,\,\,\,\,\,\,\,\,\,\,\,\,\,\,\,\,\,\,\,\,\,\,\,\,\,\,\,\,\,\,\,\,\,\,\,\,\,\,\,\,\,\,\,\,\,\, - {\delta ^2}\frac{{\left( {24\ln 2 + {\pi ^2}} \right){\gamma _2} + 48{\gamma _1}\ln 2\left( {2 + \ln 2} \right)}}{{48R_i^5{\gamma _1}}} + O\left( {{\delta ^{{5 \mathord{\left/
 {\vphantom {5 2}} \right.
 \kern-\nulldelimiterspace} 2}}}} \right).
\end{array}
\end{equation}
Comparing the different terms in (\ref{eq39}) and (\ref{eq314}) leads to the following relations:
\begin{equation} \label{eq315}
{B_0} = {R_i},\,\,\,\,\,\,{B_1} = 0,\,\,\,\,\,\,{B_2} = R_i^{ - 2}\ln 2,\,\,\,\,\,\,{B_3} = 0.
\end{equation}
Next, in order to formulate a uniform, composite solution, the inner and outer expressions are added, followed by subtracting the terms which are mutual to both of them when $\tau  \gg 1$, utilizing the relations in (\ref{eq315}). After returning to the original parameters using the relation $\delta  \buildrel \Delta \over = C_2^{ - 1}$, the latter yields the following solution, approximating the motion of the system in (\ref{eq223}) throughout deflation, in close proximity to a stable equilibrium:
\begin{equation}\label{eq316}
\begin{array}{l}
{R_{deflation}}\left( T \right) = {R_i} - C_2^{ - 1}\frac{{{\gamma _2}\left[ {\tanh \left( {{R_i}\sqrt {{\gamma _1}{C_2}} T} \right) + 2} \right]{T^2} + 4{R_i}\ln \left( {\cosh \left( {{R_i}\sqrt {{\gamma _1}{C_2}} T} \right)} \right)}}{{4R_i^3}}\\
 - C_2^{ - {3 \mathord{\left/
 {\vphantom {3 2}} \right.
 \kern-\nulldelimiterspace} 2}}\frac{{3\left[ {4{\gamma _1} + {\gamma _2}\left( {1 - 2\ln 2} \right)} \right]\tanh \left( {{R_i}\sqrt {{\gamma _1}{C_2}} T} \right)T + 2R_i^3{\gamma _1}{\gamma _3}{T^3}}}{{12R_i^4\sqrt {{\gamma _1}} }}\\
 - C_2^{ - 2}\left[ \begin{array}{l}
{\gamma _2}\frac{{\left( {12L{i_2}\left( { - {{\rm{e}}^{ - 2{R_i}\sqrt {{\gamma _1}{C_2}} T}}} \right) + {\pi ^2}} \right)\tanh \left( {{R_i}\sqrt {{\gamma _1}{C_2}} T} \right) + 2{\pi ^2}}}{{48R_i^5{\gamma _1}}}\\
 + {\gamma _2}\frac{{L{i_2}\left( { - {{\rm{e}}^{ - 2{R_i}\sqrt {{\gamma _1}{C_2}} T}}} \right) + L{i_2}\left( { - {{\rm{e}}^{2{R_i}\sqrt {{\gamma _1}{C_2}} T}}} \right) - 2\ln \left( {\cosh \left( {{R_i}\sqrt {{\gamma _1}{C_2}} T} \right)} \right)}}{{4R_i^5{\gamma _1}}}\\
 + \frac{{\ln \left( {\cosh \left( {{R_i}\sqrt {{\gamma _1}{C_2}} T} \right)} \right)\left[ {\ln \left( {\cosh \left( {{R_i}\sqrt {{\gamma _1}{C_2}} T} \right)} \right) - 2} \right]}}{{R_i^5}}
\end{array} \right] + O\left( {{\delta ^{{5 \mathord{\left/
 {\vphantom {5 2}} \right.
 \kern-\nulldelimiterspace} 2}}}} \right),
\end{array}
\end{equation}
where ${\gamma _3}$ equals to ${\beta _3}$, after substituting ${B_0} = {R_i}$.

Finally, since the solution given by (\ref{eq316}) describes only the deflation dynamics of the system, it should be supplemented by a formulation, capturing the behaviour throughout inflation. Combination of the two expressions yields a piecewise, uniform solution, depicting the behaviour of the system throughout both inflation and deflation. Namely, the behaviour of the system can be described uniformly, by switching between the solutions, describing its inflation and deflation, where the transitions occur when the time derivative of the $R$ crosses zero. Therefore, in order to complete the formulation of the deflation dynamics, the transition time ${T_{d \to i}}$, which assumes that the system starts deflating from ${R_i}$ at $T = 0$, is calculated. This is done by nulling the time derivative of (\ref{eq316}) under the assumption that $\tau  \gg 1$,  and solving the obtained quadratic algebraic equation, what leads to the following expression:
\begin{equation}\label{eq317}
{\left( {{T_{d \to i}}} \right)_{1,2}} =  - C_2^{{1 \mathord{\left/
 {\vphantom {1 2}} \right.
 \kern-\nulldelimiterspace} 2}}\frac{{4{\gamma _1} + {\gamma _2} \pm \sqrt {{{\left( {4{\gamma _1} + {\gamma _2}} \right)}^2} - 8R_i^4{\gamma _1}{\gamma _3} + 2\left( {1 + 2\ln 2} \right)C_2^{ - 1}{R_i}{\gamma _3}\left( {4{\gamma _1} + {\gamma _2}} \right)} }}{{2R_i^2\sqrt {{\gamma _1}} {\gamma _3}}}.
\end{equation}
Here, since \({T_{d \to i}}\) should be positive, the plus sign in front of the square root of the discriminant is taken.

The next stage of the derivation deals with approximating the behaviour of the second equation in (\ref{eq223}), describing the dynamics of the system throughout inflation. Since the only time derivative of $R$ included in this equation originates in a linear inertial term, it can be integrated by quadratures \citep[][]{manevich2011tractable}. However, for sake of simplicity, and since the desired approximation should describe the dynamics of the system locally, near a stable equilibrium radius, a more convenient approach, approximating the solution based on the linearized equation, is taken. This is possible since the only term causing nonlinearity in the inflation equation emanates from the elasticity of the balloon, illustrated in figure \ref{Figure7}, which can be approximated as linear in close proximity of each stable equilibrium radius. Thus, the desired approximated solution is obtained by solving the linear, leading order equation, achieved by developing this solution into a first order asymptotic series around a designated, stable equilibrium radius ${R_{eq}}$. Namely, substituting \({R_{inflation}}\left( T \right) = {R_{eq}} + \delta {R_1}\left( T \right) + O\left( {{\delta ^2}} \right)\) into the inflation equation, and solving the expression, consisting all the terms of order \(\delta \) with the initial conditions given by ${R_{inflation}}\left( {T = 0} \right) = {R_i},\,\,\,\,\,\,{\left. {{{{\rm{d}}{R_{inflation}}} \mathord{\left/
 {\vphantom {{{\rm{d}}{R_{inflation}}} {{\rm{d}}T}}} \right.
 \kern-\nulldelimiterspace} {{\rm{d}}T}}} \right|_{T = 0}} = 0$, yields a particular solution for ${R_1}$. Substituting this solution into the asymptotic series leads to the following first order approximation, describing the free motion of the system throughout inflation: 
\begin{equation}\label{eq318}
{R_{inflation}}\left( T \right) = {R_{eq}} + \left( {{R_i} - {R_{eq}}} \right)\cos \left( {\alpha T} \right) + O\left( {{\delta ^2}} \right),
\end{equation}
where after utilizing the static relation (\ref{eq31}), the constant expression of $\alpha $ is given by \(\alpha  = \sqrt {{C_3}\left[ {{S_1}\left( { - 1 + 7R_{eq}^{ - 6}} \right) + {S_2}\left( {R_{eq}^2 + 5R_{eq}^{ - 4}} \right)} \right]} \). One should notice that imposing the initial conditions cancels the small parameter $\delta$ from the coefficient of the first order function, implying that this solution is valid only if \(\left| {{R_i} - {R_{eq}}} \right| = O\left( \delta  \right)\). 

Recalling that in order to combine the inflation and deflation solutions into a uniform approximation, the transition time where the system switches from inflation to deflation should be formulated. From (\ref{eq318}) it is clear that in this case, the transition time which assumes that the system starts inflating from ${R_i}$ at $T = 0$ is \({T_{i \to d}} = \pi {\alpha ^{ - 1}}\).

\begin{figure}
\centering
\includegraphics[width=0.65\textwidth]{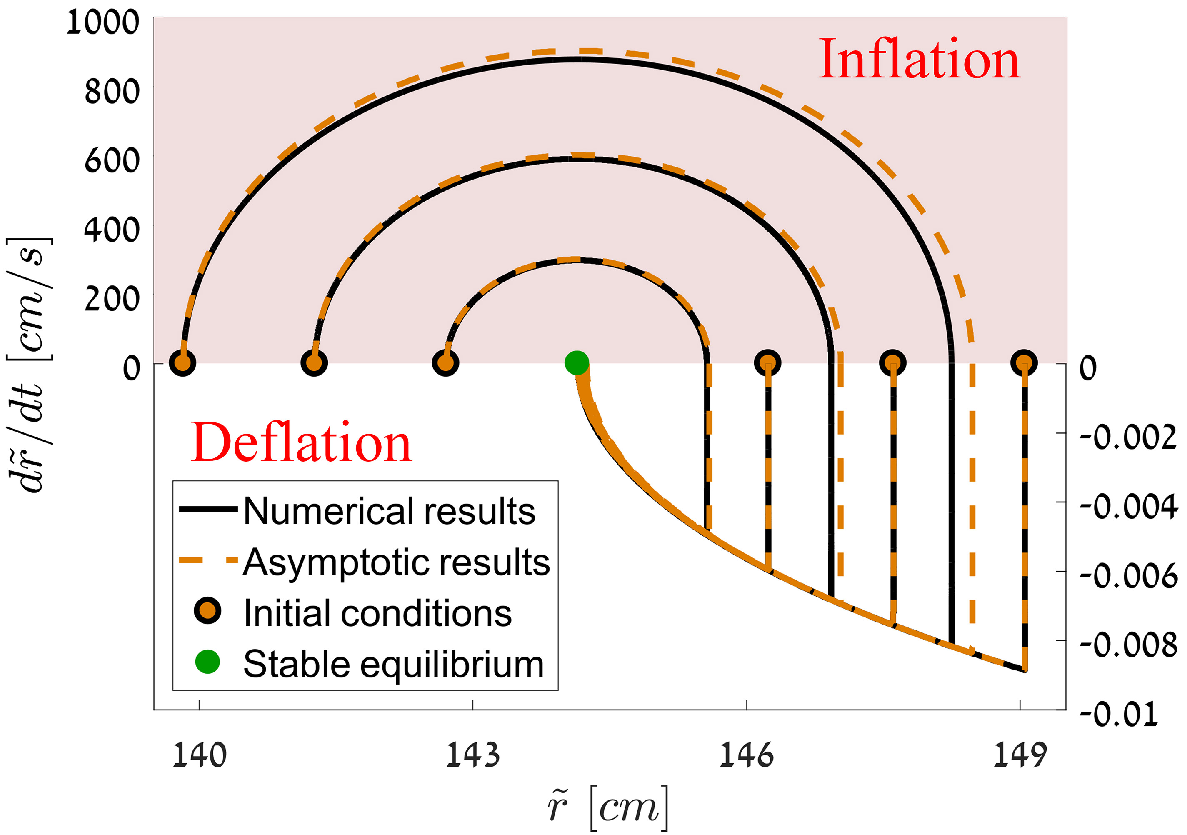}\\
\includegraphics[width=0.65\textwidth]{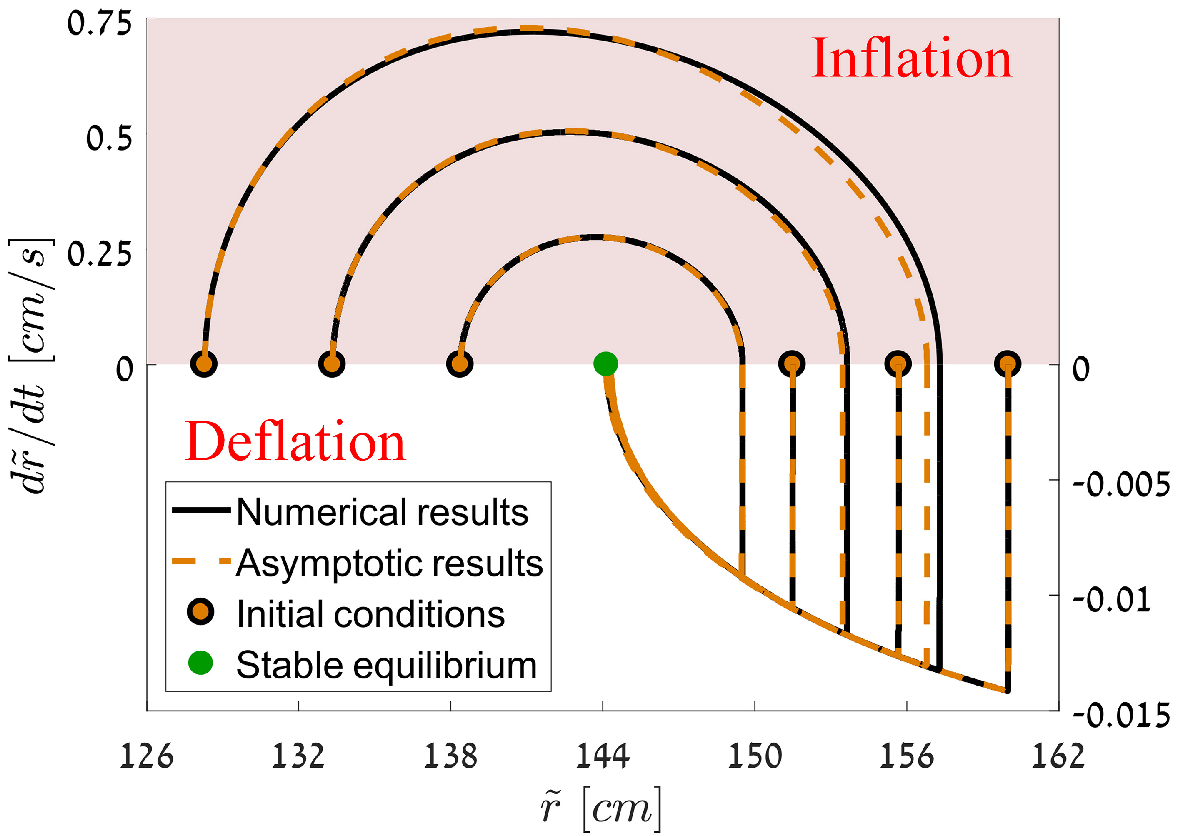}
\caption{Numerically simulated responses (solid black curves) and their corresponding asymptotic solutions (dashed orange curves), on the state-space, around the larger stable equilibrium radius (green dot), where the initial conditions are represented by the black and orange dots. The top figure describes the systems in (\ref{eq223}) and bottom figure corresponds to the system in (\ref{eq225}).}
\label{Figure8}
\end{figure}

Figure \ref{Figure8} top compares between several responses, generated by solving (\ref{eq223}) numerically, and their corresponding approximated forms, achieved utilizing the asymptotic solutions (\ref{eq316}), (\ref{eq318}). Here, the chosen geometric and physical parameters are the same as in section \ref{sec25}, where the radius of the channel is taken as ${r_{ch}} = 3\left[ {cm} \right]$, and the nominal pressure ${P_0}$ is determined as the average value between the local extrema of the static pressure-radius relation, see figure \ref{Figure7}. It should be noted that all the responses given in figure \ref{Figure8} are in close proximity to the larger stable equilibrium radius since in this region the stiffness is closer to linearity, which is consistent with the assumption that led to (\ref{eq318}). 

Figure \ref{Figure8} shows the three distinct behaviours of the system, discussed above. These behaviours include inflation, corresponding to the trajectories on the top half plane, fast deflation given by the nearly vertical curves on the bottom half plane, and slow deflation given by the curve on the bottom half plane, to which all the responses converge. It should be noted that following the slow deflation, the system keeps oscillating around the stable equilibrium radius, in amplitudes which are significantly smaller compared to the scale of interest here. These oscillations are captured to some extent by the suggested approximated solution. However, the oscillatory regime can be described in a more precise and simple manner by replacing the suggested solution of the deflation dynamics, by a second order approximation, based on an analysis, considering small oscillations around the equilibrium. Next, it can be seen from figure \ref{Figure8} that the only noticeable deviations occur throughout inflation, where these deviations increase as the initial condition gets farther from the equilibrium. The latter is due to the assumption made in the derivation of (\ref{eq318}), enforcing the initial conditions to be close to the equilibrium. On the other hand, initial conditions which are too close to the stable equilibrium lead to deviations throughout deflation since as abovementioned, a more suitable analysis for this region should rely on small amplitudes around the equilibrium, rather than on large values of ${C_2}$.

\subsubsection{Long channel}
In similar to the previous section, to approximate the free motion of the system in (\ref{eq225}), its deflation and inflation equations are solved separately, where the approximated solutions are utilized alternately each time the time derivative of the balloon's radius, crosses zero. Fortunately, the equations describing the deflation dynamics of both systems are identical, meaning that the approximated solution (\ref{eq316}) and transition time (\ref{eq317}), describe the deflation dynamics in this case as well, where ${C_1},\,{C_2}$ are replaced by ${\tilde C_1},\,{\tilde C_2}$. However, since the equation, describing the inflation of the system in (\ref{eq225}) contain a nonlinear term of the degree of freedom's time derivative, a first order approximation should not suffice in this case. Therefore, the approximated solution, describing the free inflation of the system in (\ref{eq225}), in close proximity to a stable equilibrium radius ${R_{eq}}$, uses a second order asymptotic series around this radius, given by \({R_{inflation}}\left( T \right) = {R_{eq}} + \delta {R_1}\left( T \right) + {\delta ^2}{R_2}\left( T \right) + O\left( {{\delta ^3}} \right)\). Substituting the suggested solution into the inflation equation, and consecutively solving the first and second order equations, together with the initial conditions ${R_{inflation}}\left( {T = 0} \right) = {R_i},\,\,\,\,\,\,{\left. {{{{\rm{d}}{R_{inflation}}} \mathord{\left/
 {\vphantom {{{\rm{d}}{R_{inflation}}} {{\rm{d}}T}}} \right.
 \kern-\nulldelimiterspace} {{\rm{d}}T}}} \right|_{T = 0}} = 0$, lead to the particular solutions of ${R_1}$ and ${R_2}$. These solutions complete the second order asymptotic approximation of the inflation dynamics, which is given by:

\begin{eqnarray}\label{eq319}
{R_{inflation}}\left( T \right) & = & {R_{eq}} + \left( {{R_i} - {R_{eq}}} \right)\cos \left( {{{\tilde \alpha }_1}T} \right) \nonumber\\ & + &  \frac{{{{\left( {{R_i} - {R_{eq}}} \right)}^2}}}{{6{R_{eq}}}}\left[ \begin{array}{l}
3\left( {7 - {{\tilde \alpha }_2}} \right) - 2\left( {5 - {{\tilde \alpha }_2}} \right)\cos \left( {{{\tilde \alpha }_1}T} \right)\\
 - \left( {11 - {{\tilde \alpha }_2}} \right)\cos \left( {2{{\tilde \alpha }_1}T} \right)
\end{array} \right] + O\left( {{\delta ^3}} \right),
\end{eqnarray}

where after utilizing the static relation (\ref{eq31}), the constants ${\tilde \alpha _1},\,{\tilde \alpha _2}$ are given by
\begin{subequations}\label{eq320}
\begin{equation}
{{\tilde \alpha }_1} = \sqrt {{{\tilde C}_1}\left[ {{S_1}\left( { - R_{eq}^{ - 4} + 7R_{eq}^{ - 10}} \right) + {S_2}\left( {R_{eq}^{ - 2} + 5R_{eq}^{ - 8}} \right)} \right]} ,
\end{equation}
\begin{equation}
{{\tilde \alpha }_2} = \tilde \alpha _1^{ - 2}{{\tilde C}_1}\left[ {3{S_1}\left( { - 2R_{eq}^{ - 4} + 7R_{eq}^{ - 10}} \right) + {S_2}\left( {7R_{eq}^{ - 2} + 20R_{eq}^{ - 8}} \right)} \right].
\end{equation}
\end{subequations}
As in the derivation of the inflation dynamics of the previous system, imposing the initial conditions eliminates the small parameter  $\delta $ from the coefficients of the first and second order expressions. Thus, once again, the suggested approximation is valid only if \(\left| {{R_i} - {R_{eq}}} \right| = O\left( \delta  \right)\).

Finally, the transition time from inflation to deflation is calculated by nulling the time derivative of (\ref{eq319}), leading to an infinite number of solutions. Here, under the assumption, requiring that \(\left| {{R_i} - {R_{eq}}} \right| \ll 1\),  the transition time is taken as \({T_{i \to d}} = \pi \tilde \alpha _1^{ - 1}\), which is the smallest real-valued, positive solution.

Figure \ref{Figure8} bottom shows a similar analysis to the one presented in the previous section, comparing between several responses achieved by solving (\ref{eq225}) numerically, and the corresponding asymptotic solutions. Once again, the geometric and physical parameters are taken as in section \ref{sec25}, where the supplementary parameters and operation conditions are taken as in section \ref{sec331}. As seen from figure \ref{Figure8}, qualitatively the systems in (\ref{eq223}) and (\ref{eq225}) behave similarly in close proximity to a stable equilibrium point. However, as already seen in figure \ref{Figure6}, the presence of the channel in this case compensates on the strong asymmetry between the magnitudes of the deflation and inflation velocities, exhibited by the system in (\ref{eq223}). The latter is since the prevailing inertial force in (\ref{eq225}), emanating from the flow inside the channel, is significantly larger compared to governing inertial force of the former system throughout inflation, originating in the inertia of the solid. According to (\ref{eq319}) the velocities throughout inflation can be further reduced by increasing the length of the channel and decreasing its radius. Finally, figure \ref{Figure8} shows that due to the higher order approximation of the inflation dynamics in this case, the asymptotic solution suggested for the system in (\ref{eq225}) is valid for a broader range of initial conditions.  Nevertheless, in similar to the previous system, initial conditions that are too close to the equilibrium radius, lead to deviations throughout deflation. Once again, a more suitable analysis, considering small oscillations around the equilibrium can improve the agreement throughout deflation in this region.

\subsection{Global unforced dynamics}
The current section deals with the global unforced behaviours of the systems in (\ref{eq223}) and (\ref{eq225}), under a constant input pressure, denoted ${P_0}$. The analyses presented here are carried out graphically, by examining the phase portraits of the systems.

\subsubsection{No channel}
Since the small parameter of the system in (\ref{eq223}) is defined by $\delta  \buildrel \Delta \over = C_2^{ - 1}$, the highest time derivative of $R$ is not included in the two leading orders of the equation describing the deflation of this system, see section \ref{sec331}. The latter implies that after a fast initial transient, the deflation can be described by the following, first order ordinary differential equation:
\begin{equation}\label{eq321}
\frac{{{\rm{d}}R}}{{{\rm{d}}T}} =  - \sqrt {{C_1}C_2^{ - 1}\left[ {{S_1}\left( {{R^{ - 5}} - {R^{ - 11}}} \right) + {S_2}\left( {{R^{ - 3}} - {R^{ - 9}}} \right) - {R^{ - 4}}{P_0}} \right]} ,
\end{equation}
where the second time derivative is omitted. This degenerated equation serves as manifolds on the phase portrait, capturing all the slowly varying deflation dynamics of the system in (\ref{eq223}). Namely, after an initial transient, the system slides on one of these manifolds, until convergence to a small-amplitude oscillatory motion around a stable equilibrium, which is once again ignored. As a result, in order to describe only the large-scale slow deflation dynamics of (\ref{eq223}), a single initial condition is sufficient. It is clear from (\ref{eq321}) that some values of $R$ can lead to complex expressions, implying that at these radii slow deflation is prohibited.

Next, since the equation, describing the system throughout inflation is conservative, in order to examine its global behaviour, it is multiplied by ${{{\rm{d}}R} \mathord{\left/
 {\vphantom {{{\rm{d}}R} {{\rm{d}}T}}} \right.
 \kern-\nulldelimiterspace} {{\rm{d}}T}}$, followed by integration with respect to time, what lead to the following conservation equation:
\begin{equation}\label{eq322}
{\left( {\frac{{{\rm{d}}R}}{{{\rm{d}}T}}} \right)^2} + {C_3}\frac{{3{S_1}\left( {2{R^2} + {R^{ - 4}}} \right) + 3{S_2}\left( {{R^4} + 2{R^{ - 2}}} \right) - 4{R^3}{P_0}}}{6} = Const.
\end{equation}
This expression, together with valid values of the right-hand side, which can be balanced by appropriate radii and their positive time derivatives, serve as trajectories describing the system throughout inflation.

\begin{figure}
\centering
\includegraphics[width=0.65\textwidth]{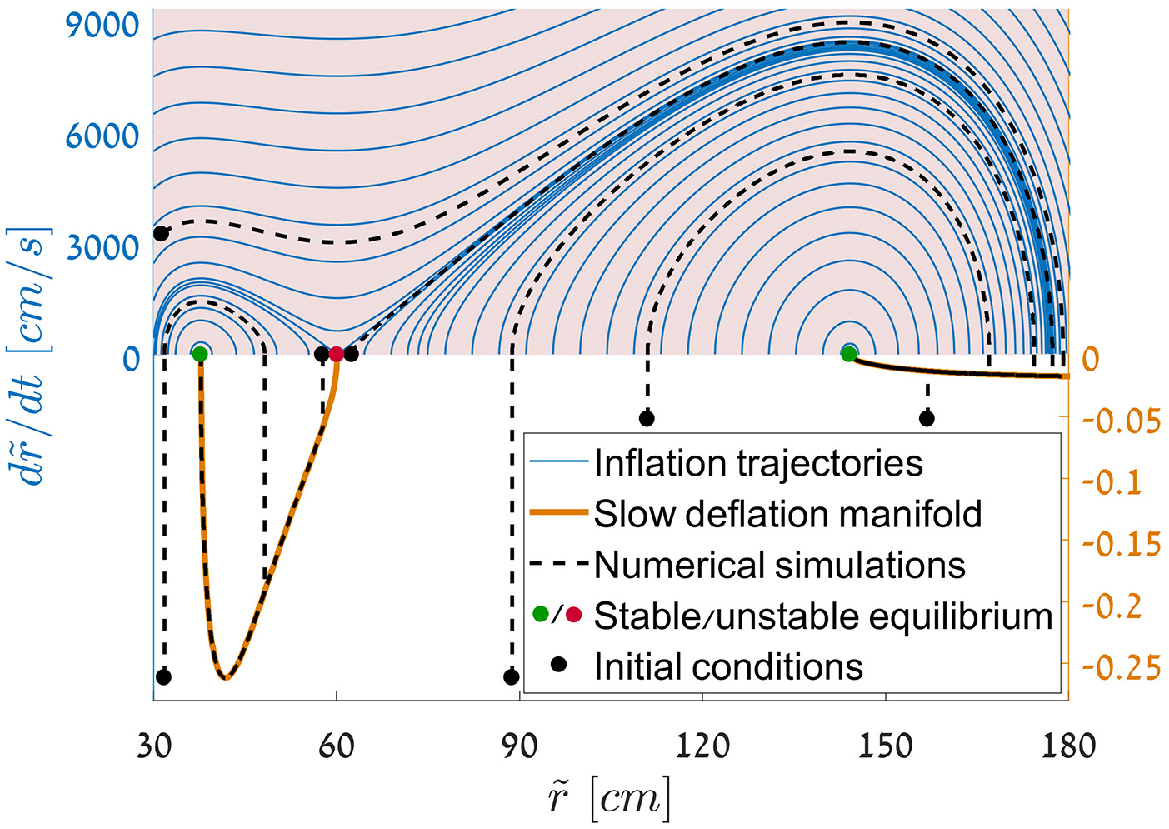}\\
\includegraphics[width=0.65\textwidth]{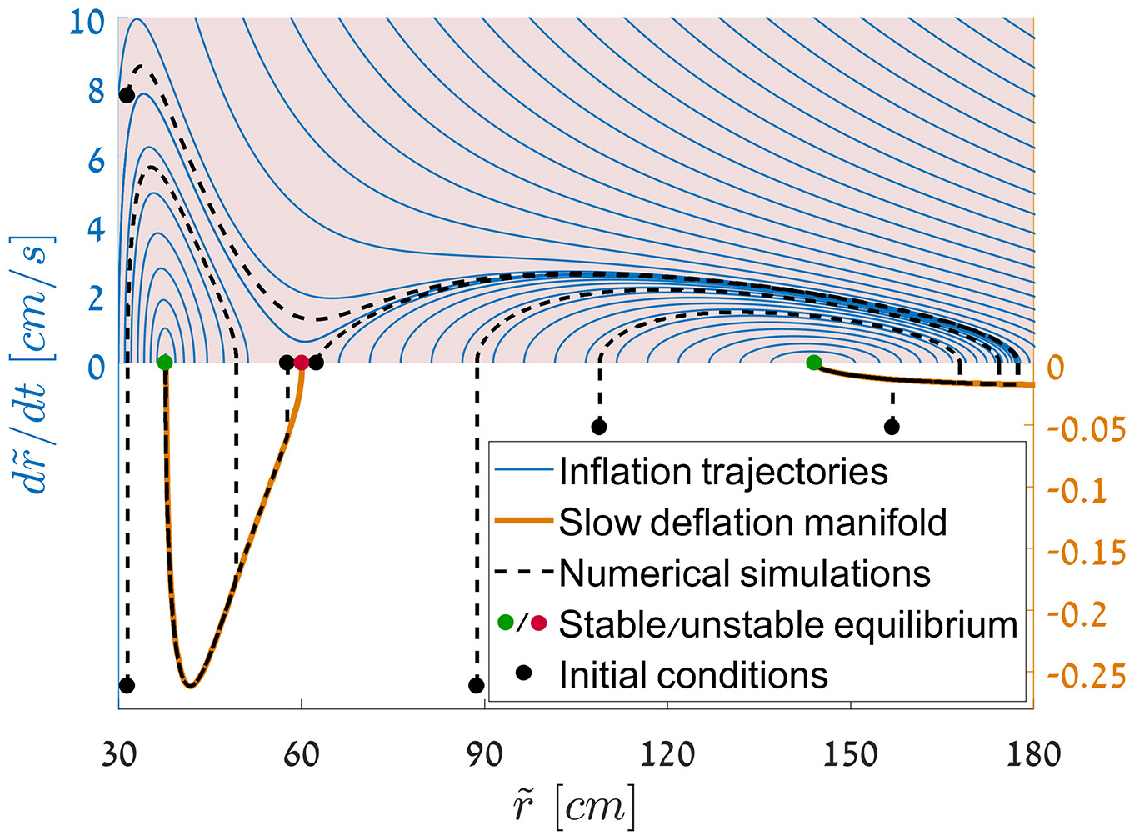}
\caption{Typical phase portraits of the systems in (\ref{eq223}) -- Top, and (\ref{eq225}) -- Bottom, in their bi-stability region. Solid blue curves -- Trajectories, describing the motion of the system throughout inflation; Solid orange curves -- The manifolds on which the system slides throughout deflation, after a rapid transient; Dashed black curves -- Simulated dynamic responses, whose initial conditions are given by the black dots; Green and red dots -- Stable and unstable equilibria, respectively.}
\label{Figure9}
\end{figure}

Figure \ref{Figure9} top presents a typical phase portrait of the system in (\ref{eq223}), in its bi-stability region, generated using the parameters and operation  conditions utilized in section \ref{sec33}. This figure shows the Isoenergetic inflation trajectories achieved using (\ref{eq322}), alongside the slow deflation manifold (\ref{eq321}), and several dynamic responses calculated by numerically solving (\ref{eq223}) from different initial conditions. From figure \ref{Figure9} it can be seen once again that the asymmetry between the two flow regimes is very strong, as the velocities related to inflation are significantly higher compared to those related to the slow deflation. Furthermore, deflation velocities which deviate from the slow manifolds (\ref{eq321}), in the intervals where they exist, converge almost instantly to the these manifolds. Similarly, in the regions where there is no slow manifold, initial conditions corresponding to deflation lead to an instant transition to inflation. Finally, since the slow deflation behaves according to a first order ordinary differential equation, once the system converges into one of the corresponding manifolds, its damping becomes significant. Thus, it relatively slowly slides on this manifold to the vicinity of a stable equilibrium, where it oscillates in very small amplitudes until convergence.

\subsubsection{Long channel}
Throughout deflation, both degenerated systems investigated in the scope of this paper behave similarly. Thus, the slow deflation of system in (\ref{eq225}) is represented by (\ref{eq321}), after replacing ${C_1},\,{C_2}$ with ${\tilde C_1},\,{\tilde C_2}$. However, the addition of the channel in this case changes the equation, describing the inflation of the system, to a form which cannot be simplified. Thus, in the typical phase portrait of the system, generated using the parameters and operation conditions utilized in section \ref{sec33}, the trajectories describing the inflation dynamics, are illustrated directly by numerically solving (\ref{eq225}) for different initial conditions. This phase portrait, illustrated in figure \ref{Figure9} bottom, shows a similar behaviour compared to the system in (\ref{eq223}), yet the inflation trajectories differ from those achieved from (\ref{eq223}), in both shape and magnitude. The latter emanates from the different governing inertial terms of the systems and from the centripetal force in the inflation equation of (\ref{eq225}), which strengthens at larger balloon diameters. Finally, it can be seen from (\ref{eq321}) and (\ref{eq226}) that the length of the channel affects the dynamics of the system throughout inflation, without influencing its behaviour throughout slow deflation. Consequentially, ${L_{ch}}$ can serve as an effective design parameter, allowing to govern only the inflation dynamics of the system.

\section{Concluding remarks}
We studied the fluid-structure interaction between a spherical hyperelastic balloon and the entrapped incompressible fluid throughout inflation and deflation, utilizing a simplified analytical model. The derivation of this model was divided into several stages, starting with formulation of the pressure applied on the balloon assuming it stays approximetly spherical throughout its motion, while distinguishing between inflation and deflation. Next, a semi-analytical model generalization algorithm was executed, based on a set of relatively simple finite-element simulations where the motion of the balloon is assumed dictated. This semi-analytical algorithm corrected the expression, describing the pressure applied on the balloon, based on an empirically obtained flow velocity profile, at the balloon's inlet. Further, the pressure field was coupled to a variational formulation of the hyperelastic balloon, yielding a non-linear, hybrid ordinary differential equation, describing the fully coupled dynamics of the system. To validate the proposed model, it was then compared to a complete finite-element scheme, describing the fluid-structure interaction of the system under investigation, with minimal simplifying assumptions. This finite-element scheme allows to examine the validity of the assumptions made during the model derivation, with an emphasis on the assumption that the balloon remains approximately spherical. Finally, the last part of the paper deals with analytical and graphical analyses of the suggested model, examining its static and unforced dynamic behaviours. The abovementioned analyses provide approximations of the system's stable equilibrium radii, a global outlook of the system's behaviour under a constant  input pressure, and an asymptotic local solution around a stable equilibrium, combining the dynamics throughout both inflation and deflation.

The analysis presented in this work yielded a simplified model capturing the fully coupled dynamics of   liquid-filled hyperelastic balloons. This model lay the foundation for the analysis of liquid-filled hyperelastic balloons in many engineering applications, including medical devices and soft robots. The presented model can be extended in future studies to examine hyperelastic balloons as building blocks for multistable systems with minimal actuation, as well as leveraging bi-stability to achieve fast reaction times.

\appendix
\section{The pressure distribution inside the balloon}\label{appA}
The expression, describing the pressure field inside the balloon, which is omitted from section \ref{sec212} is given as following:
\begin{equation}\label{eqA1}
\begin{array}{l}
p\left( {r,\theta ,t} \right) = {p_{ext}} - \frac{{{\rho _f}{L_{ch}}\left( {2 - \varepsilon } \right)}}{{\varepsilon \left( {1 - \varepsilon } \right)}}\left[ {\frac{{{{\rm{d}}^2}\tilde r}}{{{\rm{d}}{t^2}}} + \frac{{2 - 4\varepsilon  + {\varepsilon ^2}}}{{\tilde r{{\left( {1 - \varepsilon } \right)}^2}}}{{\left( {\frac{{{\rm{d}}\tilde r}}{{{\rm{d}}t}}} \right)}^2}} \right]\\
 - {\rho _f}\left[ {1 + \frac{{r\cos \theta }}{{\tilde r\left( {1 - \varepsilon } \right)}} + \sum\limits_{m = 1}^\infty  {\frac{{{A_m}}}{{2m}}\left( {{{\left( {\frac{r}{{\tilde r}}} \right)}^m}{P_m}\left( {\cos \theta } \right) - {{\left( {\varepsilon  - 1} \right)}^m}} \right)} } \right]\tilde r\frac{{{{\rm{d}}^2}\tilde r}}{{{\rm{d}}{t^2}}}\\
 + {\rho _f}\left[ \begin{array}{l}
\frac{{\varepsilon \left( {2 - \varepsilon } \right)}}{{{{\left( {1 - \varepsilon } \right)}^2}}}\left( {\frac{{r\cos \theta }}{{\tilde r\left( {1 - \varepsilon } \right)}} + 1} \right)\\
 + \sum\limits_{m = 1}^\infty  {\frac{1}{{2m}}\left[ {\frac{{{\rm{d}}{A_m}}}{{{\rm{d}}\varepsilon }}\frac{{\varepsilon \left( {2 - \varepsilon } \right)}}{{1 - \varepsilon }} + {A_m}\left( {m - 1} \right)} \right]\left[ {{{\left( {\frac{r}{{\tilde r}}} \right)}^m}{P_m}\left( {\cos \theta } \right) - {{\left( {\varepsilon  - 1} \right)}^m}} \right]} \\
 + \sum\limits_{m = 1}^\infty  {\sum\limits_{n = 1}^\infty  {\frac{{{A_m}{A_n}}}{{8mn}}\left[ \begin{array}{l}
{\left( {1 - \varepsilon } \right)^{m + n - 2}}\left[ {\mathop {\lim }\limits_{\theta  \to \pi } {B_{mn}} + mn{{\left( { - 1} \right)}^{m + n}}} \right]\\
 - {\left( {\frac{r}{{\tilde r}}} \right)^{m + n - 2}}\left[ {{B_{mn}}\left( \theta  \right) + {C_{mn}}\left( \theta  \right)} \right]
\end{array} \right]} } 
\end{array} \right]{\left( {\frac{{{\rm{d}}\tilde r}}{{{\rm{d}}t}}} \right)^2}.
\end{array}
\end{equation}
Here, the functions ${B_{mn}}\left( \theta  \right),\,\,{C_{mn}}\left( \theta  \right)$ are defined by 
\begin{subequations}\label{eqA2}
\begin{equation}
{B_{mn}}\left( \theta  \right) = \frac{1}{{4{{\sin }^2}\theta }}\left[ \begin{array}{l}
\left( {\left( {2m + 1} \right)\cos 2\theta  + 1} \right){P_m}\left( {\cos \theta } \right)\\
 - 2\left( {m + 1} \right)\cos \theta {P_{m + 1}}\left( {\cos \theta } \right)
\end{array} \right]\left[ \begin{array}{l}
\left( {\left( {2n + 1} \right)\cos 2\theta  + 1} \right){P_n}\left( {\cos \theta } \right)\\
 - 2\left( {n + 1} \right)\cos \theta {P_{n + 1}}\left( {\cos \theta } \right)
\end{array} \right],
\end{equation}
\begin{equation}
{C_{mn}}\left( \theta  \right) = \left[ \begin{array}{l}
\left( {2m + 1} \right)\cos \theta {P_m}\left( {\cos \theta } \right)\\
 - \left( {m + 1} \right){P_{m + 1}}\left( {\cos \theta } \right)
\end{array} \right]\left[ \begin{array}{l}
\left( {2n + 1} \right)\cos \theta {P_n}\left( {\cos \theta } \right)\\
 - \left( {n + 1} \right){P_{n + 1}}\left( {\cos \theta } \right)
\end{array} \right],
\end{equation}
\end{subequations}
where it can be shown numerically that \(\mathop {\lim }\limits_{\theta  \to \pi } {B_{mn}} = 0\), thus it is eliminated.

\section{The functions, describing the generalized force applied by the entrapped fluid throughout deflation}\label{appB}
The functions ${g_1}\left( \varepsilon  \right) \div {g_5}\left( \varepsilon  \right)$, which are omitted from the expression, describing the generalized force acting by the entrapped fluid throughout deflation in section \ref{sec212}, are given by:
\begin{subequations}\label{eqB}
\begin{equation}
{g_1}\left( \varepsilon  \right) = 2\pi \left( {2 - \varepsilon } \right),
\end{equation}
\begin{equation}
{g_2}\left( \varepsilon  \right) = \frac{{2\pi {{\left( {2 - \varepsilon } \right)}^2}}}{{\varepsilon \left( {1 - \varepsilon } \right)}},
\end{equation}
\begin{equation}
{g_3}\left( \varepsilon  \right) = \pi \left[ {\frac{{{{\left( {2 - \varepsilon } \right)}^2}}}{{1 - \varepsilon }} + \sum\limits_{m = 1}^\infty  {\frac{{{A_m}}}{m}\left( {\frac{{{P_{m - 1}}\left( {\varepsilon  - 1} \right) - {P_{m + 1}}\left( {\varepsilon  - 1} \right)}}{{2m + 1}} - \left( {2 - \varepsilon } \right){{\left( {\varepsilon  - 1} \right)}^m}} \right)} } \right],
\end{equation}
\begin{equation}
{g_4}\left( \varepsilon  \right) =  - \frac{{2\pi {{\left( {2 - \varepsilon } \right)}^2}\left( {2 - 4\varepsilon  + {\varepsilon ^2}} \right)}}{{\varepsilon {{\left( {1 - \varepsilon } \right)}^3}}},
\end{equation}
\begin{equation}
{g_5}\left( \varepsilon  \right) = 2\pi \left[ \begin{array}{l}
\frac{{\varepsilon {{\left( {2 - \varepsilon } \right)}^3}}}{{2{{\left( {1 - \varepsilon } \right)}^3}}} + \sum\limits_{m = 1}^\infty  {\frac{1}{{2m}}\left[ {\frac{{{\rm{d}}{A_m}}}{{{\rm{d}}\varepsilon }}\frac{{\varepsilon \left( {2 - \varepsilon } \right)}}{{1 - \varepsilon }} + {A_m}\left( {m - 1} \right)} \right]\left[ \begin{array}{l}
\frac{{{P_{m - 1}}\left( {\varepsilon  - 1} \right) - {P_{m + 1}}\left( {\varepsilon  - 1} \right)}}{{2m + 1}}\\
 - \left( {2 - \varepsilon } \right){\left( {\varepsilon  - 1} \right)^m}
\end{array} \right]} \\
 + \sum\limits_{m = 1}^\infty  {\sum\limits_{n = 1}^\infty  {\frac{{{A_m}{A_n}}}{8}\left[ \begin{array}{l}
\left( {2 - \varepsilon } \right){\left( {\varepsilon  - 1} \right)^{m + n - 2}}\\
 - \frac{1}{{mn}}\int\limits_0^{\pi  - {{\sin }^{ - 1}}\sqrt {\varepsilon \left( {2 - \varepsilon } \right)} } {\sin \theta \left( {{B_{mn}} + {C_{mn}}} \right){\rm{d}}\theta } 
\end{array} \right]} } 
\end{array} \right].
\end{equation}
\end{subequations}

\bibliographystyle{jfm}
\bibliography{Manuscript}

\begin{thebibliography}{27}
\expandafter\ifx\csname natexlab\endcsname\relax\def\natexlab#1{#1}\fi
\def\au#1{#1} \def\ed#1{#1} \def\yr#1{#1}\def\at#1{#1}\def\jt#1{\textit{#1}}
  \def\bt#1{#1}\def\bvol#1{\textbf{#1}} \def\vol#1{#1} \def\pg#1{#1}
  \def\publ#1{#1}\def\arxiv#1{#1}\def\org#1{#1}\def\st#1{\textit{#1}}

\bibitem[Ben-Haim {\em et~al.\/}(2019)Ben-Haim, Salem, Or \&
  Gat]{ben2019single}
{\sc \au{Ben-Haim, Eran}, \au{Salem, Lior}, \au{Or, Yizhar} \& \au{Gat,
  Amir~D}} \yr{2019}  \at{Single-input control of multiple fluid-driven elastic
  actuators via interaction between bi-stability and viscosity}.  \jt{arXiv
  preprint arXiv:1903.04280} .

\bibitem[Cathey~Jr(2009)]{cathey2009nasa}
{\sc \au{Cathey~Jr, HM}} \yr{2009}  \at{The nasa super pressure balloon--a path
  to flight}.  \jt{Advances in Space Research}  \bvol{44}~(1),  \pg{23--38}.

\bibitem[Gamus {\em et~al.\/}(2017)Gamus, Salem, Ben-Haim, Gat \&
  Or]{gamus2017interaction}
{\sc \au{Gamus, Benny}, \au{Salem, Lior}, \au{Ben-Haim, Eran}, \au{Gat, Amir~D}
  \& \au{Or, Yizhar}} \yr{2017}  \at{Interaction between inertia, viscosity,
  and elasticity in soft robotic actuator with fluidic network}.  \jt{IEEE
  Transactions on Robotics}  \bvol{34}~(1),  \pg{81--90}.

\bibitem[G{\'e}radin \& Rixen(2014)]{geradin2014mechanical}
{\sc \au{G{\'e}radin, Michel} \& \au{Rixen, Daniel~J}} \yr{2014} {\em
  Mechanical vibrations: theory and application to structural dynamics\/}.
  \publ{John Wiley \& Sons}.

\bibitem[Glozman {\em et~al.\/}(2010)Glozman, Hassidov, Senesh \&
  Shoham]{glozman2010self}
{\sc \au{Glozman, Daniel}, \au{Hassidov, Noam}, \au{Senesh, Merav} \&
  \au{Shoham, Moshe}} \yr{2010}  \at{A self-propelled inflatable earthworm-like
  endoscope actuated by single supply line}.  \jt{IEEE Transactions on
  Biomedical Engineering}  \bvol{57}~(6),  \pg{1264--1272}.

\bibitem[Goldstein(1965)]{goldstein1965modern}
{\sc \au{Goldstein, S.}} \yr{1965} {\em Modern developments in fluid dynamics:
  an account of theory and experiment relating to boundary layers, turbulent
  motion and wakes\/}.  \publ{Dover Publications}.

\bibitem[Holmes(2012)]{holmes2012introduction}
{\sc \au{Holmes, Mark~H}} \yr{2012} {\em Introduction to perturbation
  methods\/}, ,  \vol{vol.~20}.  \publ{Springer Science \& Business Media}.

\bibitem[Kawamura {\em et~al.\/}(2008)Kawamura, Yasuda, Tanaka, Uno, Ueda,
  Sanada \& Nakajima]{kawamura2008clinical}
{\sc \au{Kawamura, Takuji}, \au{Yasuda, Kenjiro}, \au{Tanaka, Kiyohito},
  \au{Uno, Koji}, \au{Ueda, Moose}, \au{Sanada, Kasumi} \& \au{Nakajima,
  Masatsugu}} \yr{2008}  \at{Clinical evaluation of a newly developed
  single-balloon enteroscope}.  \jt{Gastrointestinal endoscopy}  \bvol{68}~(6),
   \pg{1112--1116}.

\bibitem[Manevich \& Gendelman(2011)]{manevich2011tractable}
{\sc \au{Manevich, Leonid~Isaakovich} \& \au{Gendelman, Oleg~V}} \yr{2011} {\em
  Tractable models of solid mechanics: Formulation, analysis and
  interpretation\/}.  \publ{Springer}.

\bibitem[Mangan \& Destrade(2015)]{mangan2015gent}
{\sc \au{Mangan, Robert} \& \au{Destrade, Michel}} \yr{2015}  \at{Gent models
  for the inflation of spherical balloons}.  \jt{International Journal of
  non-linear mechanics}  \bvol{68},  \pg{52--58}.

\bibitem[Maul \& Kim(1994)]{maul1994image}
{\sc \au{Maul, Christine} \& \au{Kim, Sangtae}} \yr{1994}  \at{Image systems
  for a stokeslet inside a rigid spherical container}.  \jt{Physics of Fluids}
  \bvol{6}~(6),  \pg{2221--2223}.

\bibitem[Milic-Emili {\em et~al.\/}(1964)Milic-Emili, Mead, Turner \&
  Glauser]{milic1964improved}
{\sc \au{Milic-Emili, J}, \au{Mead, JTURNERJM}, \au{Turner, JM} \& \au{Glauser,
  EM}} \yr{1964}  \at{Improved technique for estimating pleural pressure from
  esophageal balloons}.  \jt{Journal of Applied Physiology}  \bvol{19}~(2),
  \pg{207--211}.

\bibitem[Mosadegh {\em et~al.\/}(2014)Mosadegh, Polygerinos, Keplinger,
  Wennstedt, Shepherd, Gupta, Shim, Bertoldi, Walsh \&
  Whitesides]{mosadegh2014pneumatic}
{\sc \au{Mosadegh, Bobak}, \au{Polygerinos, Panagiotis}, \au{Keplinger,
  Christoph}, \au{Wennstedt, Sophia}, \au{Shepherd, Robert~F}, \au{Gupta,
  Unmukt}, \au{Shim, Jongmin}, \au{Bertoldi, Katia}, \au{Walsh, Conor~J} \&
  \au{Whitesides, George~M}} \yr{2014}  \at{Pneumatic networks for soft
  robotics that actuate rapidly}.  \jt{Advanced functional materials}
  \bvol{24}~(15),  \pg{2163--2170}.

\bibitem[M{\"u}ller \& Strehlow(2004)]{muller2004rubber}
{\sc \au{M{\"u}ller, Ingo} \& \au{Strehlow, Peter}} \yr{2004} {\em Rubber and
  rubber balloons: paradigms of thermodynamics\/}, ,  \vol{vol. 637}.
  \publ{Springer Science \& Business Media}.

\bibitem[Mungan(2011)]{mungan2011bernoulli}
{\sc \au{Mungan, Carl~E}} \yr{2011}  \at{The bernoulli equation in a moving
  reference frame}.  \jt{European Journal of Physics}  \bvol{32}~(2),
  \pg{517}.

\bibitem[Nayfeh(2008)]{nayfeh2008perturbation}
{\sc \au{Nayfeh, Ali~H}} \yr{2008} {\em Perturbation methods\/}.  \publ{John
  Wiley \& Sons}.

\bibitem[Saito {\em et~al.\/}(2014)Saito, Iijima, Matsuzaka, Matsushima,
  Tanaka, Kajiwara \& Shimadu]{saito2014development}
{\sc \au{Saito, Y}, \au{Iijima, I}, \au{Matsuzaka, Y}, \au{Matsushima, K},
  \au{Tanaka, S}, \au{Kajiwara, K} \& \au{Shimadu, S}} \yr{2014}
  \at{Development of a super-pressure balloon with a diamond-shaped net}.
  \jt{Advances in Space Research}  \bvol{54}~(8),  \pg{1525--1529}.

\bibitem[Sankar \& Norman(2009)]{sankar2009embedded}
{\sc \au{Sankar, P} \& \au{Norman, Suresh~R}} \yr{2009} Embedded system for
  monitoring atmospheric weather conditions using weather balloon.  \bt{In {\em
  2009 International Conference on Control, Automation, Communication and
  Energy Conservation\/}},  \pg{pp. 1--4}. IEEE.

\bibitem[Schlichting \& Gersten(2016)]{schlichting2016boundary}
{\sc \au{Schlichting, Hermann} \& \au{Gersten, Klaus}} \yr{2016} {\em
  Boundary-layer theory\/}.  \publ{Springer}.

\bibitem[Si{\'e}fert {\em et~al.\/}(2019)Si{\'e}fert, Reyssat, Bico \&
  Roman]{siefert2019bio}
{\sc \au{Si{\'e}fert, Emmanuel}, \au{Reyssat, Etienne}, \au{Bico, Jos{\'e}} \&
  \au{Roman, Beno{\^\i}t}} \yr{2019}  \at{Bio-inspired pneumatic shape-morphing
  elastomers}.  \jt{Nature materials}  \bvol{18}~(1),  \pg{24}.

\bibitem[Tezduyar \& Sathe(2007)]{tezduyar2007modelling}
{\sc \au{Tezduyar, Tayfun~E} \& \au{Sathe, Sunil}} \yr{2007}  \at{Modelling of
  fluid--structure interactions with the space--time finite elements: solution
  techniques}.  \jt{International Journal for Numerical Methods in Fluids}
  \bvol{54}~(6-8),  \pg{855--900}.

\bibitem[Usha \& Nigam(1993)]{usha1993flow}
{\sc \au{Usha, R} \& \au{Nigam, SD}} \yr{1993}  \at{Flow in a spherical cavity
  due to a stokeslet}.  \jt{Fluid dynamics research}  \bvol{11}~(1-2),
  \pg{75}.

\bibitem[Wang \& Sonnenblick(1979)]{wang1979dynamic}
{\sc \au{Wang, CY} \& \au{Sonnenblick, EH}} \yr{1979}  \at{Dynamic pressure
  distribution inside a spherical ventricle}.  \jt{Journal of biomechanics}
  \bvol{12}~(1),  \pg{9--12}.

\bibitem[Wang {\em et~al.\/}(2018)Wang, Xu, Huo \& Potier-Ferry]{wang2018snap}
{\sc \au{Wang, T}, \au{Xu, F}, \au{Huo, Y} \& \au{Potier-Ferry, Michel}}
  \yr{2018}  \at{Snap-through instabilities of pressurized balloons:
  Pear-shaped bifurcation and localized bulging}.  \jt{International Journal of
  Non-Linear Mechanics}  \bvol{98},  \pg{137--144}.

\bibitem[White(1994)]{white1994fluid}
{\sc \au{White, F.~M.}} \yr{1994} {\em Fluid Mechanics\/}.  \publ{McGraw-Hill}.

\bibitem[Yamamoto {\em et~al.\/}(2001)Yamamoto, Sekine, Sato, Higashizawa,
  Miyata, Iino, Ido \& Sugano]{yamamoto2001total}
{\sc \au{Yamamoto, Hironori}, \au{Sekine, Yutaka}, \au{Sato, Yukihiro},
  \au{Higashizawa, Toshihiko}, \au{Miyata, Tomohiko}, \au{Iino, Satoru},
  \au{Ido, Kenichi} \& \au{Sugano, Kentaro}} \yr{2001}  \at{Total enteroscopy
  with a nonsurgical steerable double-balloon method}.  \jt{Gastrointestinal
  endoscopy}  \bvol{53}~(2),  \pg{216--220}.

\bibitem[Yang {\em et~al.\/}(2017)Yang, He, Sun, Chen, Shi, Xu, Chen \&
  Zhou]{yang2017optimal}
{\sc \au{Yang, Yan-Lin}, \au{He, Xuan}, \au{Sun, Xiu-Mei}, \au{Chen, Han},
  \au{Shi, Zhong-Hua}, \au{Xu, Ming}, \au{Chen, Guang-Qiang} \& \au{Zhou,
  Jian-Xin}} \yr{2017}  \at{Optimal esophageal balloon volume for accurate
  estimation of pleural pressure at end-expiration and end-inspiration: an in
  vitro bench experiment}.  \jt{Intensive care medicine experimental}
  \bvol{5}~(1),  \pg{35}.

\end{thebibliography}

\end{document}